\documentclass[aps]{revtex4}
\usepackage{amssymb,epsf}
\usepackage{amsfonts}
\usepackage{amsmath}

\begin{document}

\title{BTZ dilatonic black holes coupled to Maxwell and Born-Infeld
electrodynamics}
\author{Seyed Hossein Hendi$^{1,2}$\footnote{
email address: hendi@shirazu.ac.ir}, Behzad Eslam Panah$^{1,2}$\footnote{%
email address: behzad.eslampanah@gmail.com}, Shahram Panahiyan$^{1,3}$%
\footnote{%
email address: sh.panahiyan@gmail.com} and Mokhtar
Hassaine$^{4}$\footnote{ email address:
hassaine@inst-mat.utalca.cl}} \affiliation{$^1$ Physics Department
and Biruni Observatory, College of Sciences, Shiraz
University, Shiraz 71454, Iran\\
$^2$ Research Institute for Astronomy and Astrophysics of Maragha (RIAAM),
Maragha, Iran\\
$^{3}$ Helmholtz-Institut Jena, Fr\"{o}belstieg 3, Jena D-07743, Germany\\
$^4$ Instituto de Matem\'{}atica y Fisica, Universidad de Talca, Casilla
747, Talca, Chile }

\begin{abstract}
Motivated by the string theory corrections in low energy limit of
both gauge and gravity sides, we consider three dimensional black
holes in the presence of dilatonic gravity and Born-Infeld
nonlinear electromagnetic field. We find that geometric behavior
of the solutions is similar to the one of hyperscaling violation
metric, asymptotically. We also investigate thermodynamics of the
solutions and show that the generalization to dilatonic gravity
introduces novel properties into thermodynamics of the black holes
which were absent in the Einstein gravity. Furthermore, we explore
the possibility of tuning out part of the dilatonic effects using
the Born-Infeld generalization.
\end{abstract}

\maketitle

%%%%%%%%%%%%%%%%%%%%%%%%%

\section{Introduction}

%%%%%%%%%%%%%%%%%%%%%%%%%%%%%%

One of the main interests in considering a dilaton field is the
fact that the low-energy limit of string theory precisely involves
a massless scalar dilaton field. This in turn has motivated the
scientific community to study dilaton gravity from different
viewpoints. The scalar dilaton field has significant impacts on
the casual structure as well as on the thermodynamical features of
the charged black holes. In fact, the presence of dilaton field
also affects the structure of spacetime geometry. The impacts are
sometimes effective concerning the asymptotic behavior of dilaton
solutions. In particular, it was proved that in presence of one or
two Liouville-type dilaton potentials, black hole spacetimes are
neither asymptotically flat nor (anti)-de Sitter
\cite{MignemiW,PolettiTW,Cai,LiouvilleI,LiouvilleII,LiouvilleIII,LiouvilleIV}%
. Nevertheless, in the case of three Liouville type dilaton
potentials, it is possible to construct dilatonic black hole
solutions in the background of (anti)-de Sitter (A)dS spacetime
\cite{GaoZh,ShDH}. Also, the coupling of a dilaton field with
other gauge fields may have profound effects on the resulting
solutions \cite{KoikawaY,GibbonsM,BrillH}. Dilaton fields can also
be relevant to the construction of black holes with rather
unconventional asymptotes. For example, charged Lifshitz black
holes with arbitrary dynamical exponent can be sustained by the
presence of at least two dilaton scalar fields
\cite{Tarrio:2011de}. The extension of this dilatonic model with a
nonlinear electrodynamics was considered in Refs.
\cite{nonlinA,nonlinB,nonlinC,Zangeneh:2015uwa,Kord}. Also, in
this context,
the holographic properties of black holes have been studied in Refs. \cite%
{Zangeneh1,Zangeneh2,Zangeneh3}. Recently, studies on neutron
stars in the context of dilaton gravity \cite{HendiBEN} as well as
black holes in dilaton gravity's rainbow \cite{HendiFEP,HendiPTEP}
have been done.

In the present work, we will focus on three-dimensional dilaton gravity. Our
motivations come from the fact that the discovery of the three-dimensional
black hole (BTZ) \cite{BTZ1} and lower-dimensional gravity have gained a lot
of interests in recent two decades \cite%
{BTZ2,Emparan,ParsonsR,MyungKM,Frodden,Bravo,HendiES,WuLZ,KordBTZ}. Indeed,
the reasons for studying three dimensional gravity theories are multiple.
For example, the near horizon geometry of three-dimensional solutions can
serve as a worthwhile model to investigate some conceptual questions about
the AdS/CFT correspondence \cite{Witten07}. Moreover, the BTZ solution is a
ground that offers many facets to explore. For example, the study of the BTZ
black hole has improved our knowledge on gravitational systems and their
interactions in three dimensions \cite{Witten07}. It also opens the possible
existence of gravitational Aharonov-Bohm effect due to the noncommutative
BTZ black holes \cite{AnacletoB}. The existence of specific relations
between these black holes and effective action in string theory \cite%
{Witten,LarranagaI} are motivations. Concerning the black hole
solutions that are BTZ-like, the current literatures contains a
lot of them. For example, the existence of BTZ black
holes/wormholes in the presence of
the nonlinear electrodynamics has been investigated in \cite%
{BTZnon,Yamazaki,HendiEPJCthree} or in higher dimensions \cite%
{BTZlikeI,BTZlikeII}. In addition, exact BTZ-like solutions were shown to
arise in massive gravity \cite{BTZmassive}, dilatonic gravity \cite%
{CM1A,CM1B,CM1C}, gravity's rainbow \cite{RainbowI,RainbowIII}, new massive
gravity \cite{NewmassiveI,NewmassiveII}, Lifshitz gravity \cite{Lifshitz},
massive gravity's rainbow \cite{massrain} (see also Refs. \cite%
{BTZgravI,BTZgravII,BTZgravIII2,BTZgravIII3,BTZgravVA,BTZgravVB,BTZgravVC}
for more details). Moreover, thermal aspects have been explored
where the existence of a phase transition between the BTZ black
hole and thermal AdS space is possible \cite{Myung2006}. It is
worth mentioning that three dimensional BTZ solution is also
interesting form a quantum point of view
\cite{LiuPW,Germani,FroddenGNP,Caputa,de la Fuente,Chaturvedi}.

Here, we consider dilatonic BTZ black holes coupled with linear and
nonlinear Born-Infeld electrodynamics. In addition to what we said for our
motivations in the previous paragraphs, we are going to investigate the
effects of nonlinearity on the properties of the solutions. Although
classical electrodynamics is well organized with the Maxwell equations
(accompanying to Lorentz force), some of their shortcomings motivate one to
consider nonlinear theory. One of the old successful theory of nonlinear
electrodynamics is the so called Born-Infeld \cite{BI}. This abelian theory
enjoys most of Maxwell properties and also its related electric field of a
point like charge is regular everywhere. The foundations of this theory
became firmly established when it is realized that one can obtain its
Lagrangian from a class of low energy limit of string theory \cite%
{BIStringI,BIStringII,BIStringIII,BIStringIV,BIStringV,BIStringVI}. Hoffmann
employed this type of nonlinear theory in context of Einstein gravity \cite%
{Hoffmann}. Then, different types of black holes in the presence of this
electrodynamics have been studied in Refs. \cite%
{BHBI1,BHBI2,BHBI3,BHBI4,BHBI5,BHBI6,BHBI7,BHBI8,BHBI9,BHBI10,BHBI11,BHBI12,BHBI13,BHBI14,BHBI15}%
. Based on the mentioned motivation, we discuss dilatonic black holes with
Maxwell and Born-Infeld theories in two separated sections.

%%%%%%%%%%%%%%%%%%%%%%%%%%%%%%%%%%%%%%%%%%%%%%%%%%%%%%%%%%%%%%

\section{BTZ black hole solutions in dilaton-Maxwell gravity}

%%%%%%%%%%%%%%%%%%%%%%%%%%%%%%%%%%%%%%%%%%%%%%%%%%%%%%%%%%%%%%
Here, we consider the three-dimension action given by
\begin{equation}
I=-\frac{1}{16\pi }\int_{\emph{M}}d^{3}x\sqrt{-g}\left[ R-4\left( \nabla
\Phi \right) ^{2}-V\left( \Phi \right) +L(h,\Phi )\right] ,  \label{Action}
\end{equation}%
where $\Phi $ is the dilaton field, $V\left( \Phi \right) $ is a scalar
potential, $R$ denotes the scalar curvature and
\begin{equation}
L(h,\Phi )=-e^{-4\alpha \Phi }h.
\end{equation}%
In the last expression, $h$ stands for the Maxwell invariant $h=h_{\mu \nu
}h^{\mu \nu }$ where $h_{\mu \nu
}=\partial_{\mu}A_{\nu}-\partial_{\nu}A_{\mu}$ and $\alpha $ represents the
dilaton coupling parameter. The variation of the action (\ref{Action}) with
respect to the metric tensor, the dilaton field ($\Phi $) and the gauge
field ($A_{\mu }$), yields%
\begin{eqnarray}
R_{\mu \nu } &=&4\left[ \partial _{\mu }\Phi \partial _{\nu }\Phi +\frac{1}{4%
}g_{\mu \nu }V(\Phi )\right] +2e^{-4\alpha \Phi }\left( h_{\mu \eta }h_{\nu
}^{\eta }-\frac{1}{2}g_{\mu \nu }h_{\lambda \eta }h^{\lambda \eta }\right) ,
\label{dilaton equation(I)} \\
&&  \notag \\
\nabla ^{2}\Phi &=&\frac{1}{8}\frac{\partial V\left( \Phi \right) }{\partial
\Phi }+\frac{\alpha }{2}e^{-4\alpha \Phi }h_{\lambda \eta }h^{\lambda \eta },
\label{dilaton equation(II)} \\
&&  \notag \\
0 &=&\partial _{\mu }\left( \sqrt{-g}e^{-4\alpha \Phi }h^{\mu \nu }\right) .
\label{Maxwell}
\end{eqnarray}

Since we are interested in electrically charged static solution, we consider
the following ansatz%
\begin{equation}
ds^{2}=-f(r)dt^{2}+\frac{dr^{2}}{f(r)}+r^{2}R^{2}(r)d\varphi ^{2},
\label{MetricI}
\end{equation}%
where $f(r)$ and $R(r)$ are two metric functions.Our study being
dedicated to black holes with a radial electric field, the
suitable choice of gauge potential is given by $A_{\mu }=\delta
_{\mu }^{t}A_{0}(r)$, where $A_{0}$ denotes the electric
potential. As usual the direct integration of the Maxwell equation
(\ref{Maxwell}) permits to express the electric field $E(r)$ as
\begin{equation}
E(r)=\frac{qe^{4\alpha \Phi }}{rR(r)},  \label{Ftr eq}
\end{equation}%
where $q$ is an integration constant which is related to the electric
charge. Now, the Einstein field equations (\ref{dilaton equation(I)}) can be
re-arranged as%
\begin{eqnarray}
eq_{tt} &:&\frac{1}{2}\left[ f^{\prime \prime }(r)+\left( \frac{1}{r}+\frac{%
R^{\prime }(r)}{R(r)}\right) f^{\prime }(r)\right] +V(\Phi )=0,  \label{Ett}
\\
&&  \notag \\
eq_{rr} &:&eq_{tt}+\left[ \frac{R^{\prime \prime }(r)}{R(r)}+\frac{%
2R^{\prime }(r)}{rR(r)}+4\Phi ^{\prime 2}(r)\right] f(r)=0,  \label{rr} \\
&&  \notag \\
eq_{\theta \theta } &:&2E^{2}(r)e^{-4\alpha \Phi (r)}+\left[ \frac{R^{\prime
\prime }(r)}{R(r)}+\frac{2R^{\prime }(r)}{rR(r)}\right] f(r)+\left[ \frac{1}{%
r}+\frac{R^{\prime }(r)}{R(r)}\right] f^{\prime }(r)+V(\Phi )=0.  \label{com}
\end{eqnarray}

It is then easy to see that the substraction of Eq. (\ref{Ett}) with Eq. (%
\ref{rr}) yields the following constraint
\begin{equation}
\frac{R^{\prime \prime }(r)}{R(r)}+\frac{2R^{\prime }(r)}{rR(r)}+4\Phi
^{\prime 2}(r)=0.  \label{Wphi}
\end{equation}

In order to transform this equation into a differential equation for the
dilaton field, we use the following judicious ansatz \footnote{%
Such ansatz has been proved to be fruitful in the case of black string
solutions of dilaton-Maxwell gravity \cite{Dehghani}} for the metric function%
\begin{equation*}
R(r)=e^{2\alpha \Phi (r)}.
\end{equation*}

This in turn implies that the dilaton scalar field can be determined to be%
\begin{equation}
\Phi (r)=\frac{\gamma }{2\alpha }\ln \left( \frac{b}{r}\right) ,  \label{Phi}
\end{equation}%
where $b$\ is an arbitrary constant and for convenience, we have defined%
\begin{equation*}
\gamma =\frac{\alpha ^{2}}{\left( \alpha ^{2}+1\right) }.
\end{equation*}

In order to find analytic solutions, the Liouville-type dilation potential
is chosen as%
\begin{equation}
V(\Phi )=2\Lambda e^{4\alpha \Phi },  \label{V(Phi)}
\end{equation}%
where $\Lambda $ is a free parameter that plays the role of a cosmological
constant. Note that this kind of potential have been used in the context of
Friedman-Robertson-Walker scalar field cosmology \cite{Ozer} as well as in
the case of Maxwell-dilaton black holes \cite{Chan,Yazadjiev}. Finally, the
remaining metric function is found to be%
\begin{equation}
f(r)=\frac{2q^{2}\left( \alpha ^{2}+1\right) ^{2}}{\alpha ^{2}}-mr^{\gamma }+%
\frac{2\Lambda r^{2}\left( \alpha ^{2}+1\right) ^{2}}{\alpha ^{2}-2}\left(
\frac{b}{r}\right) ^{2\gamma },  \label{Maxf(r)}
\end{equation}%
where $m$ is an integration constant related to the mass of black holes.

Before continuing our study, it is interesting to re-write the metric
solution (\ref{MetricI}) in the "standard form" by defining the radial
coordinate $\rho =rR(r)$%
\begin{equation*}
ds^{2}=-F(\rho )\,dt^{2}+\frac{b^{\frac{2\gamma }{\gamma -1}}}{(1-\gamma
)^{2}}\frac{d\rho ^{2}}{\rho ^{\frac{2\gamma }{\gamma -1}}F(\rho )}+\rho
^{2}d\varphi ^{2},
\end{equation*}%
with%
\begin{equation*}
F(\rho )=\frac{2q^{2}\left( \alpha ^{2}+1\right) ^{2}}{\alpha ^{2}}-m\rho ^{%
\frac{\gamma }{1-\gamma }}b^{\frac{-\gamma ^{2}}{1-\gamma }}+\frac{2\Lambda
\left( \alpha ^{2}+1\right) ^{2}}{\alpha ^{2}-2}\rho ^{2}.
\end{equation*}

It is interesting to note that for $\alpha \rightarrow 0$ with $q/\alpha
\rightarrow 0$, the solution reduces to the uncharged static BTZ black hole.
Asymptotically for $\rho >>1$, the metric behaves as%
\begin{equation*}
ds^{2}\sim -\rho ^{2}dt^{2}+\frac{d\rho ^{2}}{\rho ^{2(1-\alpha ^{2})}}+\rho
^{2}d\varphi ^{2},\qquad \alpha ^{2}<2,
\end{equation*}%
or%
\begin{equation*}
ds^{2}\sim -\rho ^{\alpha ^{2}}dt^{2}+\rho ^{\alpha ^{2}}d\rho ^{2}+\rho
^{2}d\varphi ^{2},\qquad \alpha ^{2}>2.
\end{equation*}

In both cases, the asymptotic behavior is similar to the one of
the hyperscaling violation metric
\begin{equation*}
ds^{2}=\frac{1}{r^{2\theta }}\left[ -r^{2z}dt^{2}+\frac{dr^{2}}{r^{2}}%
+r^{2}d\varphi ^{2}\right] \sim -\rho ^{\frac{2(z-\theta )}{1-\theta }%
}dt^{2}+\frac{d\rho ^{2}}{\rho ^{\frac{2}{1-\theta }}}+\rho ^{2}d\varphi
^{2},
\end{equation*}%
where $z$ is the Lifshitz dynamical exponent and $\theta $ is the
hyperscaling violating parameter. More precisely, for $\alpha ^{2}<2$, the
asymptotic metric corresponds to an hyperscaling violation metric with $z=1$
and $\theta =\alpha ^{2}/(\alpha ^{2}-1)$, and for $\alpha ^{2}>2$, this
corresponds to $z=2/\alpha ^{2}$ and $\theta =(2+\alpha ^{2})/\alpha ^{2}$.

The charged dilatonic BTZ black holes are different from the charged BTZ
black holes in Einstein gravity. The existence of the dilaton may exchange
the role of the mass with the charge and \textit{vice et versa}. Indeed, in
the Einstein-Maxwell gravity, the mass is associated to the constant term of
the metric function while the charge term appears in the structural metric
function with a function depending on the radial coordinate. As one can note
from the expression (\ref{Maxf(r)}), in the dilatonic case, this is exactly
the opposite that occurs. In addition, it is worth mentioning that the
obtained charged dilatonic BTZ solution does not reduce to the charged BTZ
black hole solution in the absence of dilaton field. It is expected and
comes from the difference between polynomial functions and logarithmic one.
This behavior is the same as comparison between higher dimensional charged
black holes and three dimensional case.

We are looking for the curvature singularity, in order to confirm the black
hole interpretation of the solutions. For this purpose, we calculate the
Ricci and Kretschmann scalars. These latter are obtained as
\begin{eqnarray}
R &=&\frac{4q^{2}}{r^{2}}-\frac{m\gamma }{r^{(\alpha ^{2}+2)/(\alpha ^{2}+1)}%
}+\frac{4\Lambda \left( 2\alpha ^{2}-3\right) }{\alpha ^{2}-2}\left( \frac{b%
}{r}\right) ^{2\gamma }, \\
&&  \notag \\
R_{\alpha \beta \mu \nu }R^{\alpha \beta \mu \nu } &=&\frac{16q^{4}}{r^{4}}+%
\frac{3\gamma ^{2}m^{2}}{r^{2\left( \alpha ^{2}+2\right) /\left( \alpha
^{2}+1\right) }}+\frac{32\left( \alpha ^{4}-2\alpha ^{2}+\frac{3}{2}\right)
\Lambda ^{2}}{\left( \alpha ^{2}-2\right) ^{2}}\left( \frac{b}{r}\right)
^{4\gamma }  \notag \\
&&-\frac{32q^{2}\left( \alpha ^{2}-1\right) \Lambda b^{2\gamma }}{\left(
\alpha ^{2}-2\right) r^{2\left( 2\alpha ^{2}+1\right) /\left( \alpha
^{2}+1\right) }}-\frac{8m\gamma q^{2}}{r^{\left( 3\alpha ^{2}+4\right)
/\left( \alpha ^{2}+1\right) }}-\frac{8\gamma m\left( 2\alpha ^{2}-1\right)
\Lambda b^{2\gamma }}{\left( \alpha ^{2}-2\right) r^{\left( 3\alpha
^{2}+2\right) /\left( \alpha ^{2}+1\right) }}.
\end{eqnarray}

Calculations show that for finite values of radial coordinate, the Ricci and
Kretschmann scalars are finite. Also, for very small and very large values
of $r$, we have%
\begin{eqnarray}
\lim_{r\rightarrow 0^{+}}R &=&\infty ,  \notag \\
\lim_{r\rightarrow 0^{+}}R_{\alpha \beta \mu \nu }R^{\alpha \beta \mu \nu
}&=&\infty ,  \label{Ricci} \\
&&  \notag \\
\lim_{r\rightarrow \infty }R &\propto &\frac{4\Lambda \left( 2\alpha
^{2}-3\right) }{\alpha ^{2}-2}\left( \frac{b}{r}\right) ^{2\gamma },  \notag
\\
\lim_{r\rightarrow \infty }R_{\alpha \beta \mu \nu }R^{\alpha \beta \mu \nu
} &\propto &\frac{32\left( \alpha ^{4}-2\alpha ^{2}+\frac{3}{2}\right)
\Lambda ^{2}}{\left( \alpha ^{2}-2\right) ^{2}}\left( \frac{b}{r}\right)
^{4\gamma }.  \label{Kresch}
\end{eqnarray}

The above equation (\ref{Ricci}) confirms that there is an essential
singularity located at $r=0$, and Eq. (\ref{Kresch}) for $\alpha =0$, the
asymptotic behavior of solutions is (A)dS ($\lim_{r\rightarrow \infty
}R\propto 6\Lambda $ and $\lim_{r\rightarrow \infty }R_{\alpha \beta \mu \nu
}R^{\alpha \beta \mu \nu }\propto 12\Lambda ^{2}$), while for nonzero $%
\alpha $, the asymptotic behavior of solutions is not that of (A)dS.

\begin{figure}[tbp]
$%
\begin{array}{ccc}
\epsfxsize=5.5cm \epsffile{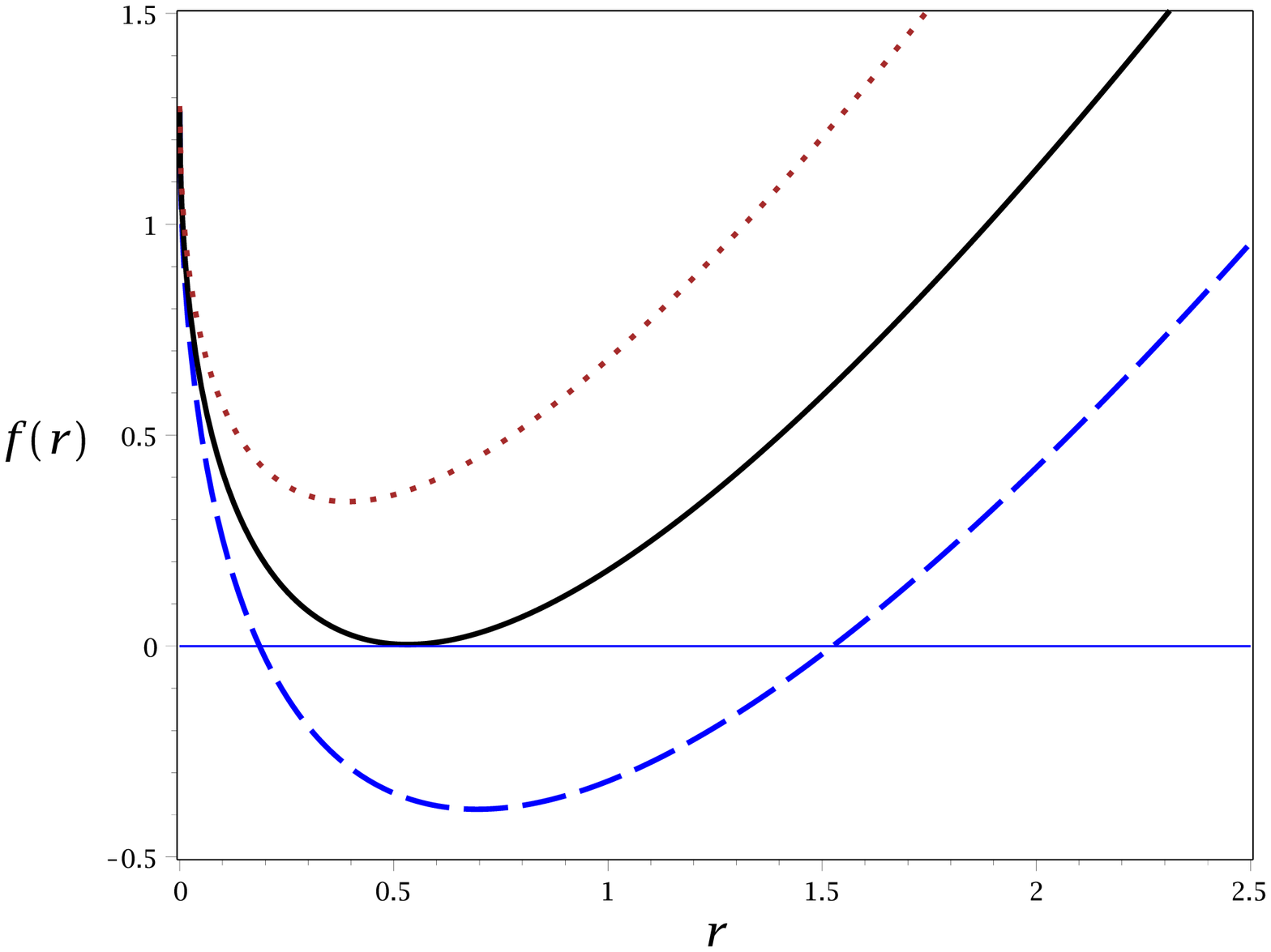} & \epsfxsize=5.5cm %
\epsffile{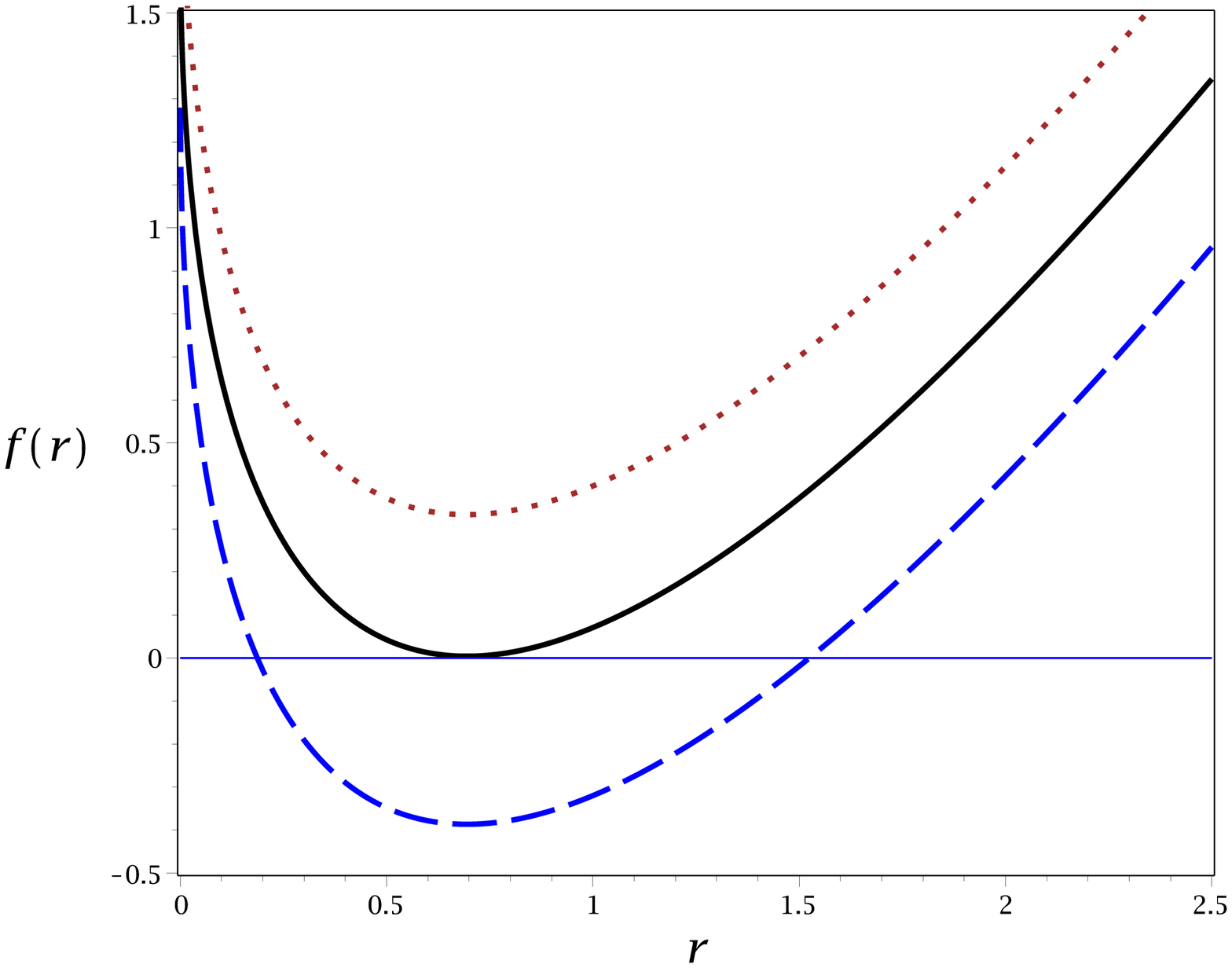} & \epsfxsize=5.5cm \epsffile{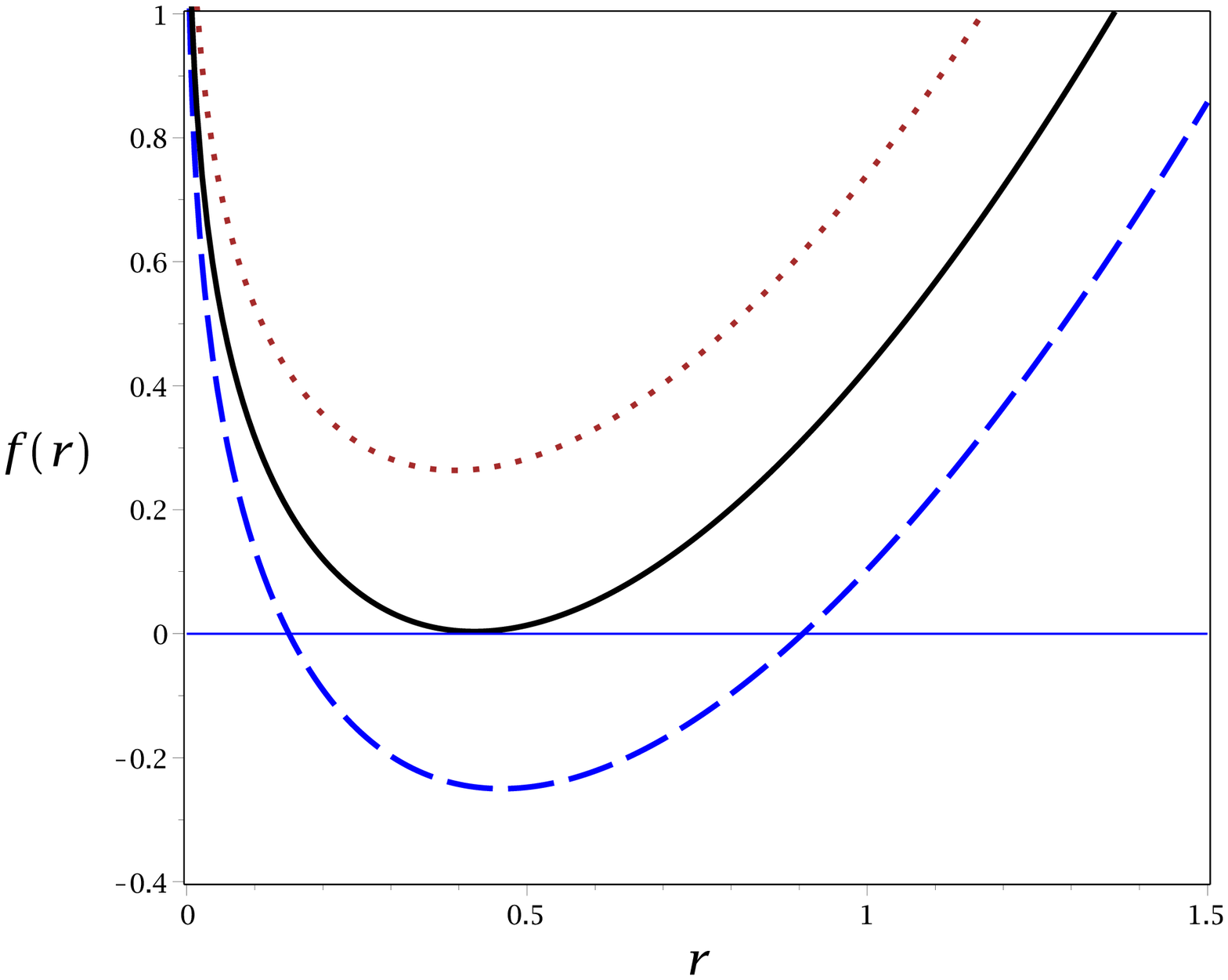}%
\end{array}
$%
\caption{$f(r)$ versus $r$, for $b=0.3$ and $\Lambda =-1$. \newline
Left panel: for $q=0.4$, $\protect\alpha =1$, $m=4$ (dashed line), $m=3.5$
(continuous line) and $m=3$ (dotted line). \newline
Middle panel: for $m=4$, $\protect\alpha =1$, $q=0.4$ (dashed line), $%
q=0.457 $ (continuous line) and $q=0.5$ (dotted line). \newline
Right panel: for $m=3$, $q=0.4$, $\protect\alpha =0.5$ (dashed line), $%
\protect\alpha =0.439$ (continuous line) and $\protect\alpha=0.4$ (dotted
line).}
\label{Fig1}
\end{figure}

For more details regarding the behavior of the metric function, we plot $%
f(r) $ versus $r$ in Fig. \ref{Fig1}. As one can see, this solution may
contain real positive roots, and therefore, the singularity can be covered
with an event horizon and interpreted as a black hole.
%%%%%%%%%%%%%%%%%%%%%%%%%%%%%%%%%%%%%%%%%%%%%%%%%%%%%%%%%%%%%%%

%%%%%%%%%%%%%%%%%%%%%%%%%%%%%%%%%%%%%%%%%%%%%%%%%%%%%%%%%%%%%%%

\subsection{Thermodynamical properties}

%%%%%%%%%%%%%%%%%%%%%%%%%%%%%%%%%%%%%%%%%%%%%%%%%%%%%%%%%%%%%%%

In this section, we are going to calculate the thermodynamic and the
conserved quantities of the solutions and then examine the first law of
thermodynamics.

In order to calculate the Hawking temperature, we employ the definition of
surface gravity. In doing so, we have that
\begin{equation}
T=\frac{1}{2\pi }\sqrt{-\frac{1}{2}\left( \nabla _{\mu }\chi _{\nu }\right)
\left( \nabla ^{\mu }\chi ^{\nu }\right) },  \label{T}
\end{equation}%
where $\chi =\partial /\partial t$ is the Killing vector. The Hawking
temperature for the this black hole can be written as
\begin{equation}
T_{+}=-\frac{\left( \alpha ^{2}+1\right) }{2\pi r_{+}}\left[ q^{2}+\Lambda
r_{+}^{2}\left( \frac{b}{r_{+}}\right) ^{2\gamma }\right] ,  \label{TemMax}
\end{equation}
where $r_{+}$ is the event horizon of black hole which is the largest real
root of metric function, that is $f(r=r_{+})=0$. In addition, the electric
potential $U$ is defined by the gauge potential in the following form \cite%
{Cvetic1}
\begin{equation}
U=A_{\mu }\chi ^{\mu }\left\vert _{r\rightarrow reference}\right. -A_{\mu
}\chi ^{\mu }\left\vert _{r=r_{+}}\right. ,  \label{U}
\end{equation}%
and for our solutions, we obtain
\begin{equation}
U=\frac{q}{\gamma }\left( \frac{b}{r_{+}}\right) ^{\gamma }.  \label{ElecBI1}
\end{equation}
Also, one can use the area law of entropy in Einstein gravity to obtain the
entropy of this black hole. Based on this law, the black hole's entropy
equals to one-quarter of horizon area \cite%
{Bekenstein1,Bekenstein2,Bekenstein3}. Therefore, the entropy is
\begin{equation}
S=\frac{\pi r_{+}}{2}\left( \frac{b}{r_{+}}\right) ^{\gamma }.
\label{entropyMax}
\end{equation}
In order to obtain the electric charge of the black holes, one can calculate
the flux of the electromagnetic field at infinity. The electric charge is
obtained as
\begin{equation}
Q=\frac{q}{2}.  \label{QMax}
\end{equation}
Regarding the timelike Killing vector $\left( \xi =\partial /\partial
t\right) $, one can show that the finite mass can be obtained as
\begin{equation}
M=\frac{m}{8}\left( 1-\gamma \right) b^{\gamma }.  \label{massMax}
\end{equation}

By evaluating the metric function on the largest root of the solution, one
is able to extract the geometrical mass ($m$) and insert into the total mass
(\ref{massMax}). This leads to following relation
\begin{equation}
M=\frac{\left( \alpha ^{2}+1\right) \Lambda r_{+}^{2}}{4\left( \alpha
^{2}-2\right) }\left( \frac{b}{r_{+}}\right) ^{3\gamma }+\frac{q^{2}\left(
\alpha ^{2}+1\right) }{4\alpha ^{2}}\left( \frac{b}{r_{+}}\right) ^{\gamma }.
\label{MMax}
\end{equation}

Using the relations obtained for the entropy (\ref{entropyMax}) and the
total electric charge (\ref{QMax}), one is able to find a Smarr-like formula
as
\begin{equation}
M(S,Q)=\frac{\left( \alpha ^{2}+1\right) \Lambda \left( \frac{2S}{\pi }%
\right) ^{2\left( \alpha ^{2}+1\right) }b^{-2\alpha ^{2}}}{4\left( \alpha
^{2}-2\right) }\left( \frac{\pi b}{2S}\right) ^{3\alpha ^{2}}+\frac{%
Q^{2}\left( \alpha ^{2}+1\right) }{\alpha ^{2}}\left( \frac{\pi b}{2S}%
\right) ^{\alpha ^{2}}.  \label{SmarrMax}
\end{equation}

Now, we can check the validity of first law of thermodynamics. In order to
achieve this task, we note that
\begin{equation}
dM(S,Q)=\left( \frac{\partial M(S,Q)}{\partial S}\right) _{Q}dS+\left( \frac{%
\partial M(S,Q)}{\partial Q}\right) _{S}dQ,
\end{equation}
and it is a matter of check to show that following equalities hold%
\begin{equation}
T=\left( \frac{\partial M}{\partial S}\right) _{Q}\ \ \ \ \&\ \ \ \ U=\left(
\frac{\partial M}{\partial Q}\right) _{S}.  \label{TU}
\end{equation}
The above relations confirm that the first law of thermodynamics is valid,
namely
\begin{equation}
dM=TdS+UdQ.
\end{equation}

%%%%%%%%%%%%%%%%%%%%%%%%%%%%%%%%%%%%%

\subsection{Thermodynamic behavior}

%%%%%%%%%%%%%%%%%%%%%%%%%%%%%%%%%%%%%
Here, the main goal in this section is to specificize the effects on the
coupling constants of the problem on the thermodynamic behavior of the
solution, particulary for the mass, the temperature and the heat capacity.

%%%%%%%%%%%%%%%%%%%%%%%%%%%%%%%%%%%%%

\subsubsection{Mass/Internal energy}

%%%%%%%%%%%%%%%%%%%%%%%%%%%%%%%%%%%%%
The mass of the black holes is usually interpreted as the internal energy of
the system. Nevertheless, in presence of a cosmological constant, the mass
can also be viewed as an enthalpy with the cosmological constant playing the
role of the thermodynamical pressure. Here, we do not consider such
possibility and instead we will regard the mass as the internal energy.

\begin{figure}[tbp]
$%
\begin{array}{cc}
\epsfxsize=7cm \epsffile{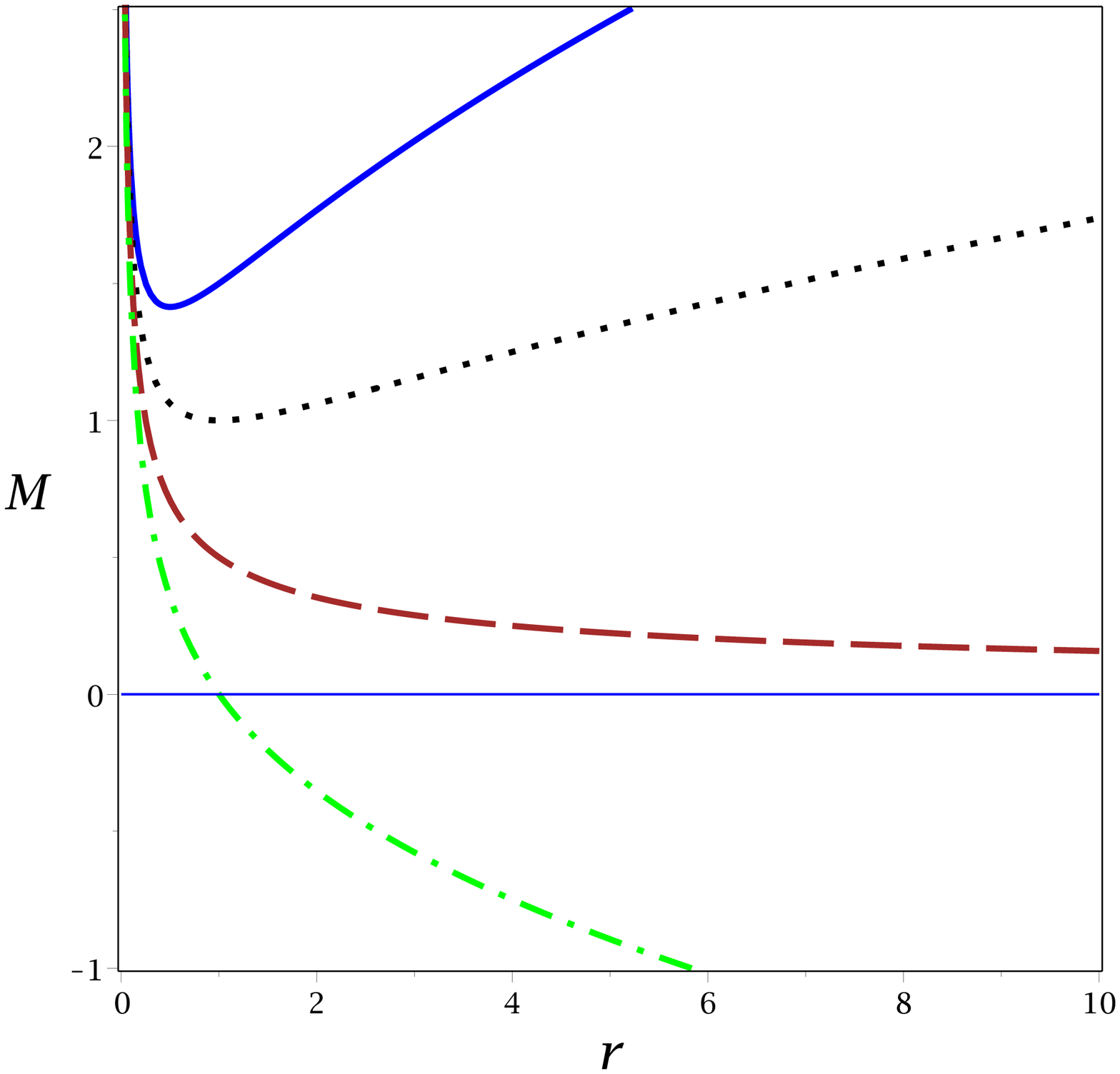} & \epsfxsize=7cm \epsffile{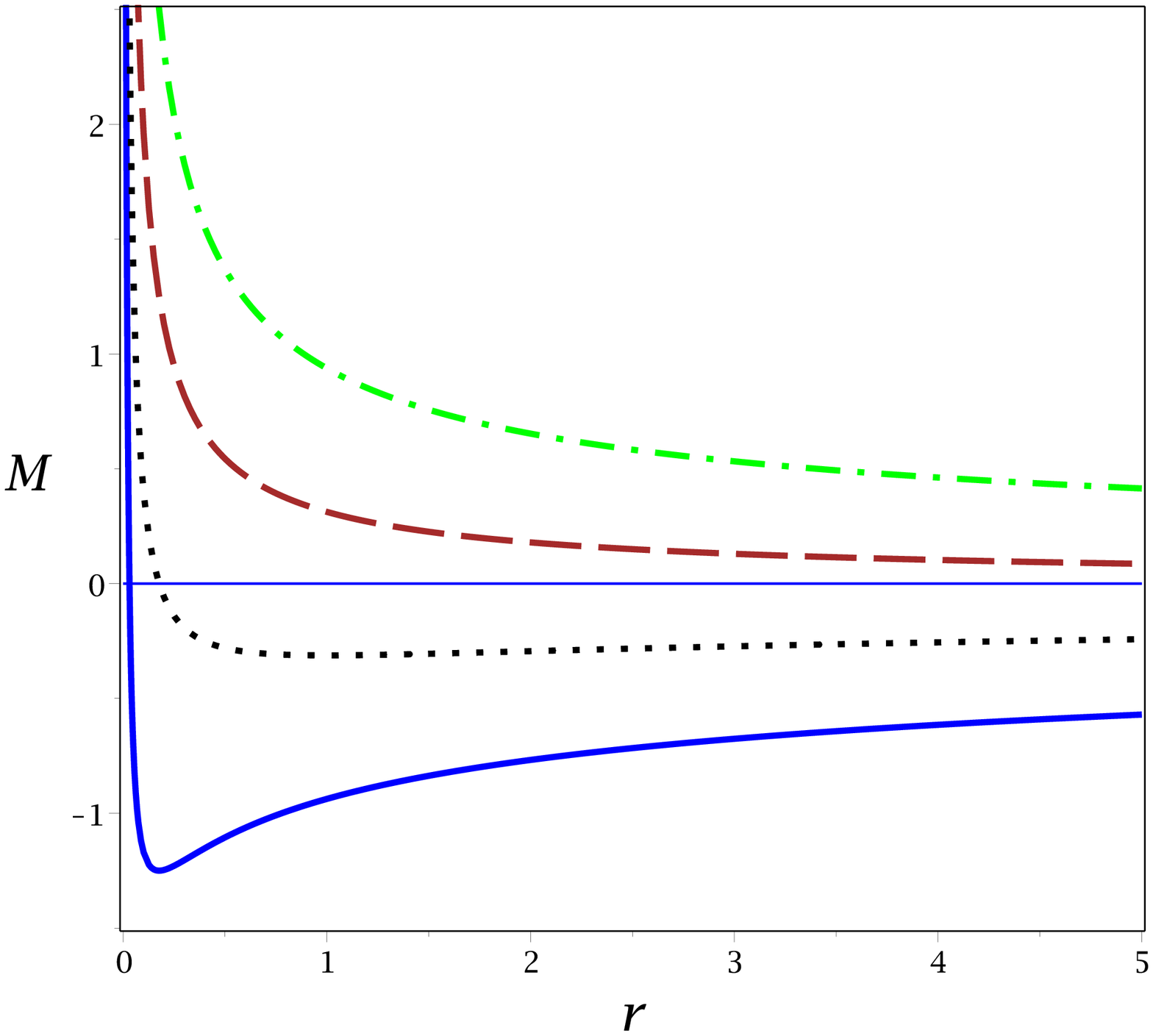}%
\end{array}
$%
\caption{$M$ versus $r_{+}$, for $b=1$, $q=1$, $\Lambda =-2$ (continuous
line), $\Lambda =-1$ (dotted line), $\Lambda =0$ (dashed line) and $\Lambda
=1$ (dashed-dotted line). \newline
Left panels: $\protect\alpha =1$; Right panels: $\protect\alpha =2$. }
\label{FigMax1}
\end{figure}

First of all, the mass as defined in (\ref{MMax}) requires the coupling
constant $\alpha\not=\pm \sqrt{2}$. Now, since the dilatonic parameter $b>0$%
, the $q^2-$ part of the mass expression is always positive. On the other
hand, it is known that the constant $\Lambda$ plays the role of a
cosmological constant in the dilaton gravity. Therefore, it could be
negative (for adS case) or positive (for dS case). Considering these two
options, one finds that the $\Lambda-$contribution of the mass expression is
positive provided that
\begin{eqnarray*}
\Lambda &>&0\text{ \ \ \ \ \ }\&\text{ \ \ \ \ \ }\alpha >\sqrt{2}, \\
\Lambda &<&0\text{ \ \ \ \ \ }\&\text{ \ \ \ \ \ }\alpha <\sqrt{2}.
\end{eqnarray*}
Under these conditions, to absence of roots and the positivity of the
internal energy are ensured. In contrast, a negative value of the $\Lambda-$%
contribution of the mass expression can yield the existence of a root and a
region of negativity for the internal energy. In such case, the root of the
internal energy is obtained as
\begin{equation}
r_{+}|_{_{M=0}} =\left( -\frac{q^{2}\left( \alpha ^{2}-2\right) }{\Lambda
\alpha ^{2}}\right) ^{\frac{\alpha ^{2}+1}{2}}b^{-\alpha ^{2}}.
\label{RootMMax}
\end{equation}
Evidently, the root of the internal energy is a decreasing function of $%
\Lambda$, while it is an increasing function of the electric charge. In
order to elaborate our results, we have plotted a series of diagrams (see
Fig. \ref{FigMax1}). Finally, considering positive internal energy as a
condition for having black holes, one can conclude that physical black hole
solutions are present in range of $0<r_{+}<r_{+}|_{_{M=0}}$. Later, we will
add some other restrictions which are imposed by temperature and heat
capacity to complete our picture for our solutions to describe physical
black holes.

%%%%%%%%%%%%%%%%%%%%%%%%%%%%%%%%%%%%%%%%%%%%%%%%%%%%%%%%%%%%%%%
%%%%%%%%%%%%%%%%%%%%%%%%%%%%%%%%%%%%%%%%%%%%%%%%%%%%%%%%%%%%%%%

\subsubsection{Temperature}

\begin{figure}[tbp]
$%
\begin{array}{cc}
\epsfxsize=7cm \epsffile{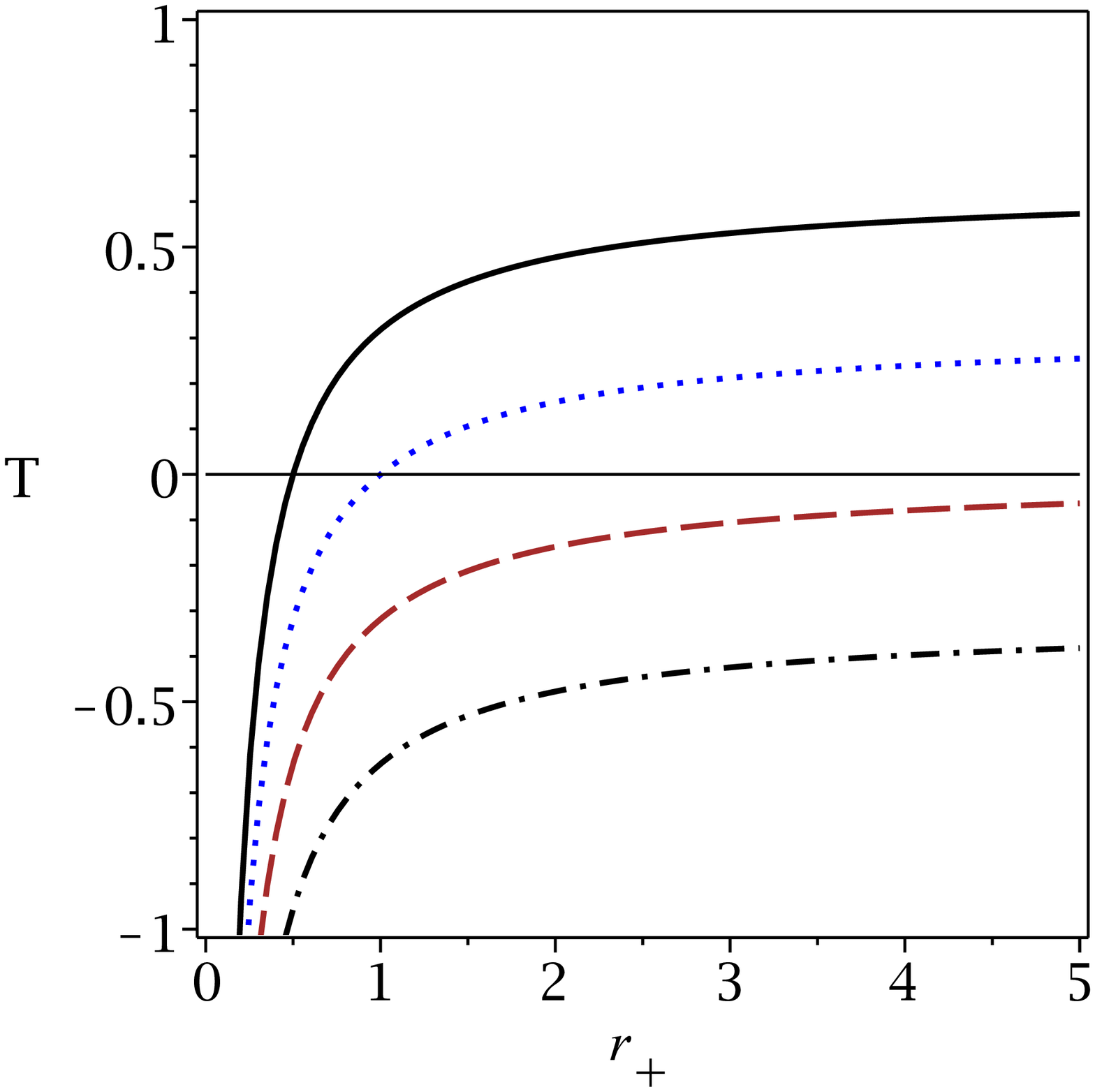} & \epsfxsize=7cm \epsffile{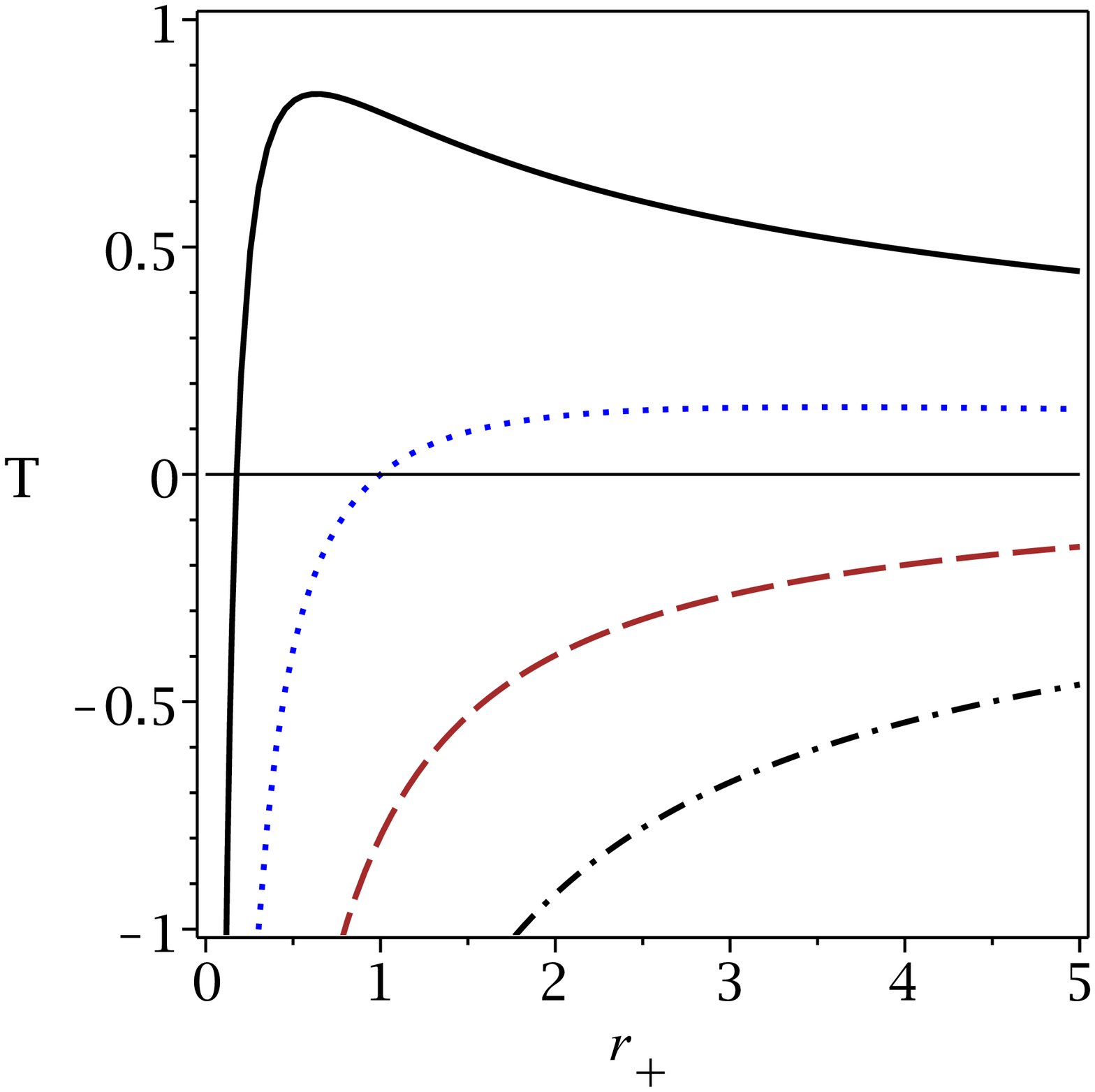}%
\end{array}
$%
\caption{$T$ versus $r_{+}$, for $b=1$, $q=1$, $\Lambda=-2$ (continuous
line), $\Lambda=-1 $ (dotted line), $\Lambda=0$ (dashed line) and $\Lambda=1$
(dashed-dotted line). \newline
Left panels: $\protect\alpha =1$; Right panels: $\protect\alpha =2$. }
\label{FigMax2}
\end{figure}

In classical thermodynamics of black holes, one of the conditions for having
physical solutions is the positivity of the temperature. This highlights the
importance of the roots of temperature. It is a matter of calculation to
show that for these black holes, we have the following root for temperature
\begin{equation}
r_{+}|_{_{T=0}} =\left( -\frac{q^{2}}{\Lambda }\right) ^{\frac{\alpha ^{2}+1%
}{2}}b^{-\alpha ^{2}}.  \label{RootTMax}
\end{equation}
Here, the root of the temperature is a decreasing function of $\Lambda$,
while it is an increasing function of the electric charge. Considering the
possibility of having both positive and negative values for $\Lambda$, the
positive valued root only exists for $\Lambda <0$. On the other hand, for
positive values of the cosmological constant ($\Lambda>0$), the temperature
will be negative. This indicates that in the classical thermodynamics of
black holes, physical solutions exist only for $\Lambda <0$ with the
following condition
\begin{equation*}
\Lambda <-\frac{q^{2}}{r_{+}^{2}}\left( \frac{b}{r_{+}}\right) ^{-2\gamma }.
\end{equation*}
Let us summarize what we have found by studying the temperature (in
classical thermodynamics of black holes). First of all, physical solutions
only exist for negative values of $\Lambda$. Also, we found an upper limit
on the values of $\Lambda$ which is obtained by the condition of having
positive temperature. Condition that bridges the values of the electric
charge, $q$ with $\Lambda$. For completeness, we also present the following
diagrams for the temperature in term of $r_+$ (see Fig. \ref{FigMax2}). In
case of the absence of root, the temperature is negative valued everywhere.
On the other hand, in presence of root, the positive valued temperature only
exists for $r_{+}|_{_{T=0}} <r_{+}$. By suitable choices of different
parameters, the temperature could also acquires an extremum (maximum). It is
a matter of calculation to show that this extremum is obtained as
\begin{equation}
r_{+}|_{_{T=T_{Maximum}}}=\left( -\frac{q^{2}\left( \alpha ^{2}+1\right) }{%
\Lambda \left( \alpha ^{2}-1\right) }\right) ^{\frac{\alpha ^{2}+1}{2}%
}b^{-\alpha ^{2}}.  \label{ExtermumTMax}
\end{equation}
Later, we will show that this extremum coincides with the divergencies of
the heat capacity.

%%%%%%%%%%%%%%%%%%%%%%%%%%%%%%%%%%%%%%%%%%%%%%%%%%%%%%%%%%%%%%%
%%%%%%%%%%%%%%%%%%%%%%%%%%%%%%%%%%%%%%%%%%%%%%%%%%%%%%%%%%%%%%%

\subsubsection{Heat Capacity}

The study of the heat capacity is of important from two perspectives.
Firstly, their discontinuities represent thermodynamical phase transition
points. Secondly, the sign of the heat capacity determines whether system is
in thermally stable/instable state.

\begin{figure}[tbp]
$%
\begin{array}{cc}
\epsfxsize=7cm \epsffile{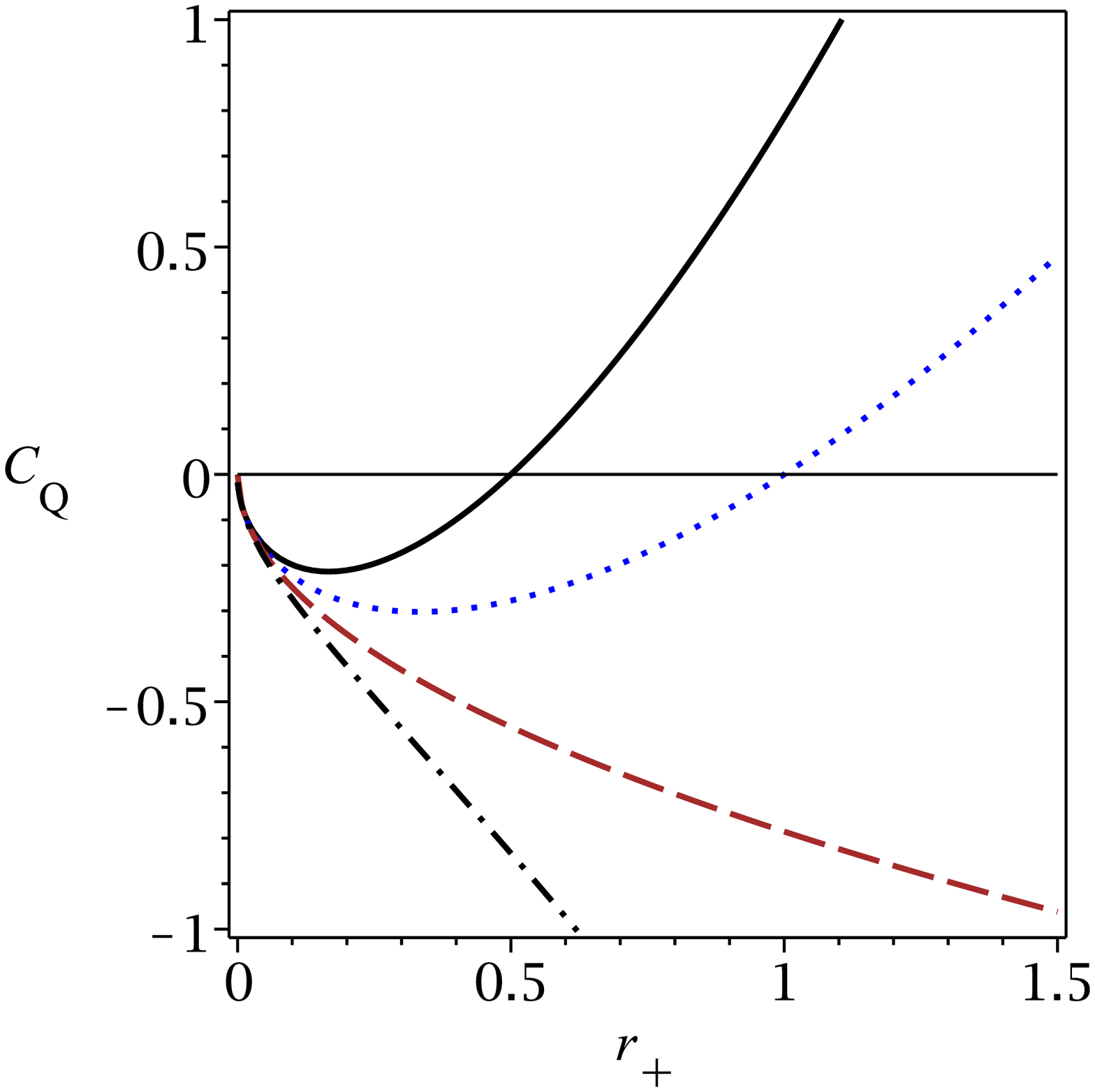} & \epsfxsize=7cm %
\epsffile{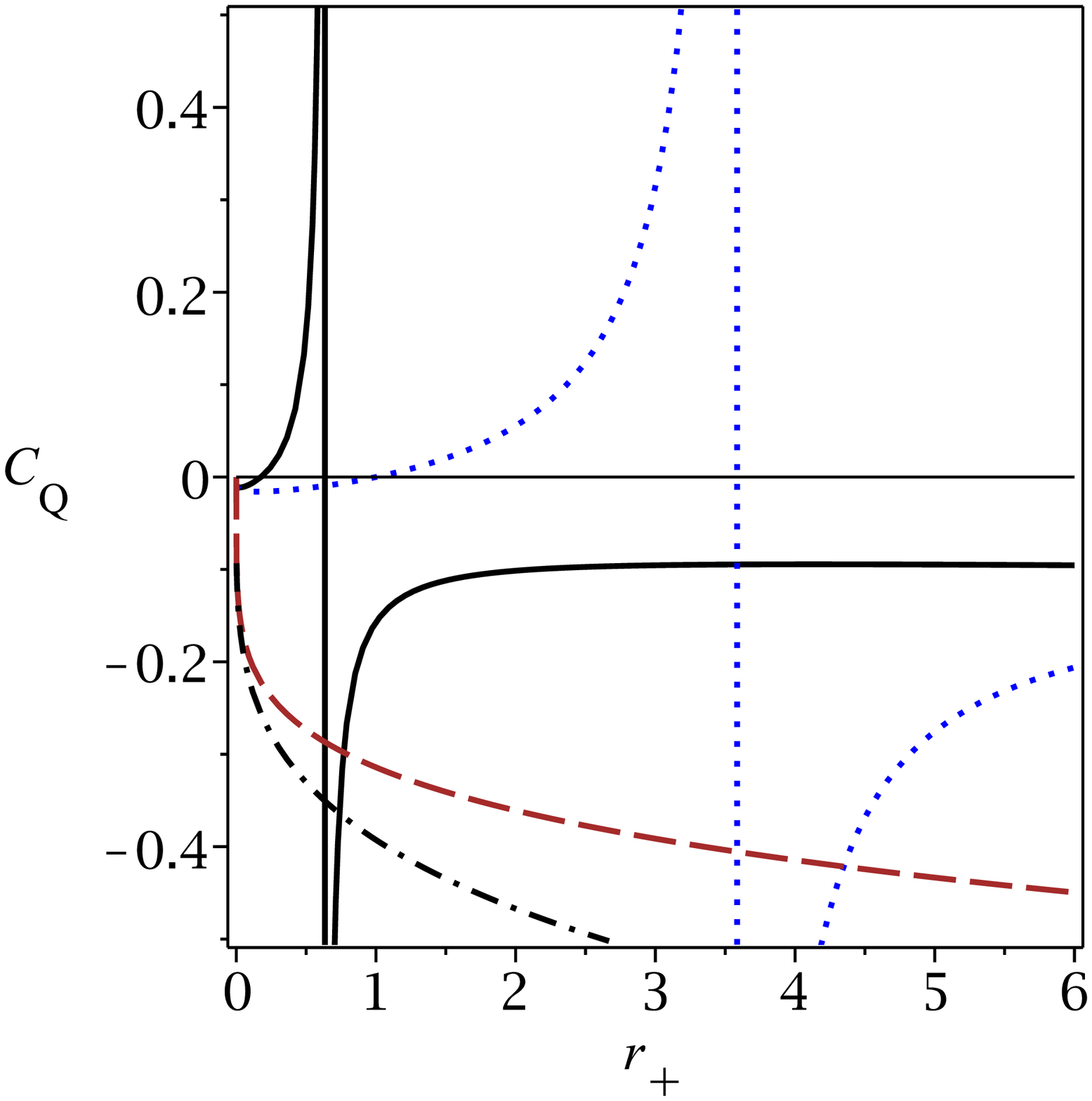}%
\end{array}
$%
\caption{$C_{Q}$ versus $r_{+}$, for $b=1$, $q=1$, $\Lambda =-2$ (continuous
line), $\Lambda =-1$ (dotted line), $\Lambda =0$ (dashed line) and $\Lambda
=1$ (dashed-dotted line). \newline
Left panels: $\protect\alpha =1$; Right panels: $\protect\alpha =2$. }
\label{FigMax3}
\end{figure}

The heat capacity is given by
\begin{equation}
C_{Q}=\frac{T}{\left( \frac{\partial ^{2}M}{\partial S^{2}}\right) _{Q}}%
=T\left( \frac{\partial S}{\partial T}\right) _{Q}=T\frac{\left( \frac{%
\partial S}{\partial r_{+}}\right) _{Q}}{\left( \frac{\partial T}{\partial
r_{+}}\right) _{Q}},  \label{Heat}
\end{equation}%
where by using Eqs. (\ref{TemMax}) and (\ref{entropyMax}), this expression
becomes
\begin{equation}
C_{Q}=-\frac{\pi r_{+}\left( \frac{b}{r_{+}}\right) ^{\gamma }\left[
q^{2}+\Lambda r_{+}^{2}\left( \frac{b}{r_{+}}\right) ^{2\gamma }\right]
\left( \alpha ^{2}+1\right) }{2q^{2}\left( 3\alpha ^{2}-1\right) -2\left(
\alpha ^{2}-1\right) \Lambda r_{+}^{2}\left( \frac{b}{r_{+}}\right)
^{2\gamma }}.  \label{HeatMax}
\end{equation}
It is a matter of calculation to show that the root and divergence points of
the heat capacity are given respectively by
\begin{eqnarray}
r_{+}\left( C_{Q}=0\right) &=&\left( -\frac{q^{2}}{\Lambda }\right) ^{\frac{%
\alpha ^{2}+1}{2}}b^{-\alpha ^{2}},  \label{RootCQMax} \\
r_{+}\left( C_{Q}\longrightarrow \infty \right) &=&\left( -\frac{q^{2}\left(
\alpha ^{2}+1\right) }{\Lambda \left( \alpha ^{2}-1\right) }\right) ^{\frac{%
\alpha ^{2}+1}{2}}b^{-\alpha ^{2}}.  \label{DivCQMax}
\end{eqnarray}
Evidently, both the heat capacity and the temperature share the same roots.
Therefore, the arguments given for the root of the temperature stand for the
heat capacity as well. In addition, the extremum for the temperature (\ref%
{ExtermumTMax}) and divergence point of heat capacity are same. In order to
have positive valued divergence point for the heat capacity, the following
conditions should be satisfied
\begin{eqnarray*}
\Lambda &>&0\text{ \ \ \ \ \ }\&\text{ \ \ \ \ \ }\alpha <1, \\
\Lambda &<&0\text{ \ \ \ \ \ }\&\text{ \ \ \ \ \ }\alpha >1.
\end{eqnarray*}
In the previous section, we have shown that only for negative values of the $%
\Lambda$, our solutions enjoy a positive temperature. Consequently, for
negative values of $\Lambda $ and $\alpha >1$, our solutions will develop a
phase transition in their thermodynamical structure. In order to have a
positive heat capacity which implies stable solutions, the denominator and
numerator of the heat capacity must be of the same sign, that is
\begin{equation*}
q+\Lambda r_{+}^{2}\left( \frac{b}{r_{+}}\right) ^{2\gamma }<0\text{ \ \ \ \
\ }\&\text{ \ \ \ \ \ }2q^{2}\left( \alpha ^{2}+1\right) +2\left( \alpha
^{2}-1\right) \Lambda r_{+}^{2}\left( \frac{b}{r_{+}}\right) ^{2\gamma }>0%
\text{ ,}
\end{equation*}%
or%
\begin{equation*}
q+\Lambda r_{+}^{2}\left( \frac{b}{r_{+}}\right) ^{2\gamma }>0\text{ \ \ \ \
\ }\&\text{ \ \ \ \ \ }2q^{2}\left( \alpha ^{2}+1\right) +2\left( \alpha
^{2}-1\right) \Lambda r_{+}^{2}\left( \frac{b}{r_{+}}\right) ^{2\gamma }<0%
\text{ .}
\end{equation*}
In order to have a better picture of the behavior of the heat capacity, we
have plotted various diagrams (see Fig. \ref{FigMax3}). It is explicitly
shown that there are three possible cases for the heat capacity; I) The heat
capacity has no root and is negative everywhere. II) The heat capacity has
only one root, in which after that, the heat capacity is positive and the
solutions are stable. III) Finally, the heat capacity enjoys one root and
one divergence point in its structure. In this case, before the root and
after the divergency, the heat capacity is negative and solutions are
thermally unstable. Whereas, only between the root and divergence point, the
system is thermally stable. To end this section, we would like to add a
comment. Previously, it was shown that the heat capacity in the context of
BTZ-$\Lambda $-Maxwell theory enjoys the existence of root only for negative
values of $\Lambda $, whereas, the divergence point was only observed for
positive values of the $\Lambda$ \cite{Mamasani}. Here, we see that the
generalization to dilaton gravity has a significant effect on such behavior.
Here, the negative values of $\Lambda $ could enjoy both root and divergency
in their structure while the branch of the positive $\Lambda $ was ruled out
due to the absence of positive valued temperature. Thermodynamically
speaking, the generalization to dilaton gravity provided black holes with
more complexity in their thermodynamical structure on level of the
introduction of new phase transition point which was absent in the previous
case. This highlights the differences between these two theories.
%%%%%%%%%%%%%%%%%%%%%%%%%%%%%%%%%%%%%%%%%%%%%%%%%%%%%%%%%%%%%%%
%%%%%%%%%%%%%%%%%%%%%%%%%%%%%%%%%%%%%%%%%%%%%%%%%%%%%%%%%%%%%%%

\section{BTZ black hole solutions in dilaton-Born-Infeld gravity}

We now turn in the derivation of the dilatonic-BI-BTZ black holes where the
Lagrangian of the BI-dilaton part is given by
\begin{equation}
L(h,\Phi )=4\beta ^{2}e^{4\alpha \Phi }\left( 1-\sqrt{1+\frac{e^{-8\alpha
\Phi }h}{2\beta ^{2}}}\right) ,  \label{LBI}
\end{equation}%
where $\beta $ is the BI parameter. It is notable that, in the limit $\beta
\rightarrow \infty $, the Lagrangian BI, reduces to the standard Maxwell
field coupled to a dilaton field as $L(h,\Phi )=-e^{4\alpha \Phi }h$.
Varying the action (\ref{Action}) with respect to the metric tensor $g_{\mu
\nu }$, the dilaton field $\Phi $ and the gauge field $A_{\mu }$, we obtain
the following field equations
\begin{eqnarray}
R_{\mu \nu } &=&4\left[ \partial _{\mu }\Phi \partial _{\nu }\Phi +\frac{1}{4%
}g_{\mu \nu }V(\Phi )\right] -4e^{-4\alpha \Phi }\partial _{Y}L\left(
Y\right) h_{\mu \eta }h_{\nu }^{\eta }+4\beta ^{2}e^{4\alpha \Phi }\left[
2Y\partial _{Y}L\left( Y\right) -L\left( Y\right) \right] g_{\mu \nu },
\label{Field equation} \\
&&  \notag \\
\nabla ^{2}\Phi &=&\frac{1}{8}\frac{\partial V\left( \Phi \right) }{\partial
\Phi }+2\alpha \beta ^{2}e^{4\alpha \Phi }\left[ 2Y\partial _{Y}L\left(
Y\right) -L\left( Y\right) \right] , \\
&&  \notag \\
0 &=&\partial _{\mu }\left( \sqrt{-g}e^{-4\alpha \Phi }\partial _{Y}L\left(
Y\right) h^{\mu \nu }\right) ,  \label{BI equation}
\end{eqnarray}%
where $L(Y)=1-\sqrt{1+Y}$ and $Y$ is given as
\begin{eqnarray}
Y &=&\frac{e^{-8\alpha \Phi }h}{2\beta ^{2}}. \\
\end{eqnarray}

In order to derive the black hole solutions of this system of equations, we
use static Ansatz metric as defined in Eq. (\ref{MetricI}) with a purely
electrically field $A_{\mu }=\delta _{\mu }^{0}A(r)$. Using the Eqs. (\ref%
{BI equation}) and (\ref{MetricI}), we have
\begin{equation}
4\beta ^{2}\left[ \alpha rE\left( r\right) R(r)\Phi ^{\prime }\left(
r\right) -\frac{\left( rE(r)R(r)\right) ^{\prime }}{4}\right] e^{8\alpha
\Phi \left( r\right) }+\left[ rR^{\prime }\left( r\right) +R(r)\right]
E^{3}\left( r\right) =0,
\end{equation}%
where we can obtain the electric field as
\begin{equation}
E(r)=\frac{dA(r)}{dr}=\frac{qe^{4\alpha \Phi }}{rR(r)\sqrt{1+\Gamma }},
\label{h(r)}
\end{equation}%
where we have defined $\Gamma =\frac{q^{2}}{r^{2}\beta ^{2}R^{2}(r)}$. For
latter convenience, we chose a Liouville-type dilation potential defined by $%
V(\Phi )=2\Lambda e^{4\alpha \Phi }$ with the ansatz $R(r)=e^{2\alpha \Phi
(r)}$. After some algebraic calculations, we obtain the following
differential equations
\begin{eqnarray}
\frac{\alpha ^{2}f(r)-r\left( 1+\alpha ^{2}\right) f^{\prime }\left(
r\right) }{r^{2}\left( 1+\alpha ^{2}\right) ^{2}}+4\left( \beta ^{2}-\frac{%
\Lambda }{2}\right) \left( \frac{b}{r}\right) ^{2\gamma }-4\beta ^{2}\left(
\frac{b}{r}\right) ^{2\gamma }\sqrt{1+\frac{q^{2}}{r^{2}\beta ^{2}\left(
\frac{b}{r}\right) ^{2\gamma }}} &=&0,  \label{tt} \\
2\left( e^{2\alpha \Phi (r)}\right) ^{\prime }+r\left( e^{2\alpha \Phi
(r)}\right) ^{\prime \prime }+4r\left( \Phi ^{\prime }\left( r\right)
\right) ^{2}e^{2\alpha \Phi (r)} &=&0.  \label{P(r)}
\end{eqnarray}

We are now in a position to obtain exact solutions. Considering the
equations (\ref{tt}) and (\ref{P(r)}), we can obtain the general solutions as%
\begin{eqnarray}
f(r) &=&\frac{2\left( \alpha ^{2}+1\right) ^{2}\left( \Lambda -2\beta
^{2}\right) r^{2}}{\alpha ^{2}-2}\left( \frac{b}{r}\right) ^{2\gamma
}-mr^{\gamma }+\frac{4\beta ^{2}\left( \alpha ^{2}+1\right) ^{2}}{\alpha
^{2}-2}\left( \frac{b}{r}\right) ^{2\gamma }r^{2}H_{1}+\frac{4q^{2}\left(
\alpha ^{2}+1\right) ^{2}}{\alpha ^{2}}H_{2},  \label{DilatonBTZBI} \\
&&  \notag \\
\Phi (r) &=&\frac{\gamma }{2\alpha }\ln \left( \frac{b}{r}\right) ,
\label{Phi(r)BI}
\end{eqnarray}%
in which\textbf{\ }$H_{1}$ and $H_{2}$ are the following hypergeometric
functions
\begin{eqnarray}
H_{1} &=&{}_2F_{1}\left( \left[ \frac{1}{2},\frac{\alpha ^{2}-2}{2}\right] ,%
\left[ \frac{\alpha ^{2}}{2}\right] ,-\Gamma \right) ,  \notag \\
&&  \notag \\
H_{2} &=&{}_2F_{1}\left( \left[ \frac{1}{2},\frac{\alpha ^{2}}{2}\right] ,%
\left[ \frac{\alpha ^{2}+2}{2}\right] ,-\Gamma \right) .  \notag
\end{eqnarray}
It is notable that, in the absence of a BI field ($\beta \rightarrow \infty $%
), the solutions (\ref{DilatonBTZBI}) reduce to the charged dilatonic BTZ
black hole solutions (see Eq. (\ref{Maxf(r)})).

Calculation of the Kretschmann scalar shows that it is finite for nonzero $r$
and its behavior for very small and very large values of $r$ can be reported
as
\begin{eqnarray}
\lim_{r\rightarrow 0^{+}}R_{\alpha \beta \mu \nu }R^{\alpha \beta \mu \nu }
&=&\infty ,  \label{RR0} \\
\lim_{r\rightarrow \infty }R_{\alpha \beta \mu \nu }R^{\alpha \beta \mu \nu
} &\propto &\frac{16\Lambda ^{2}(2\alpha ^{4}-4\alpha ^{2}+3)}{\left( \alpha
^{2}-2\right) ^{2}}\left( \frac{b}{r}\right) ^{4\gamma }.  \label{RRinf}
\end{eqnarray}

\begin{figure}[tbp]
$%
\begin{array}{cc}
\epsfxsize=7cm \epsffile{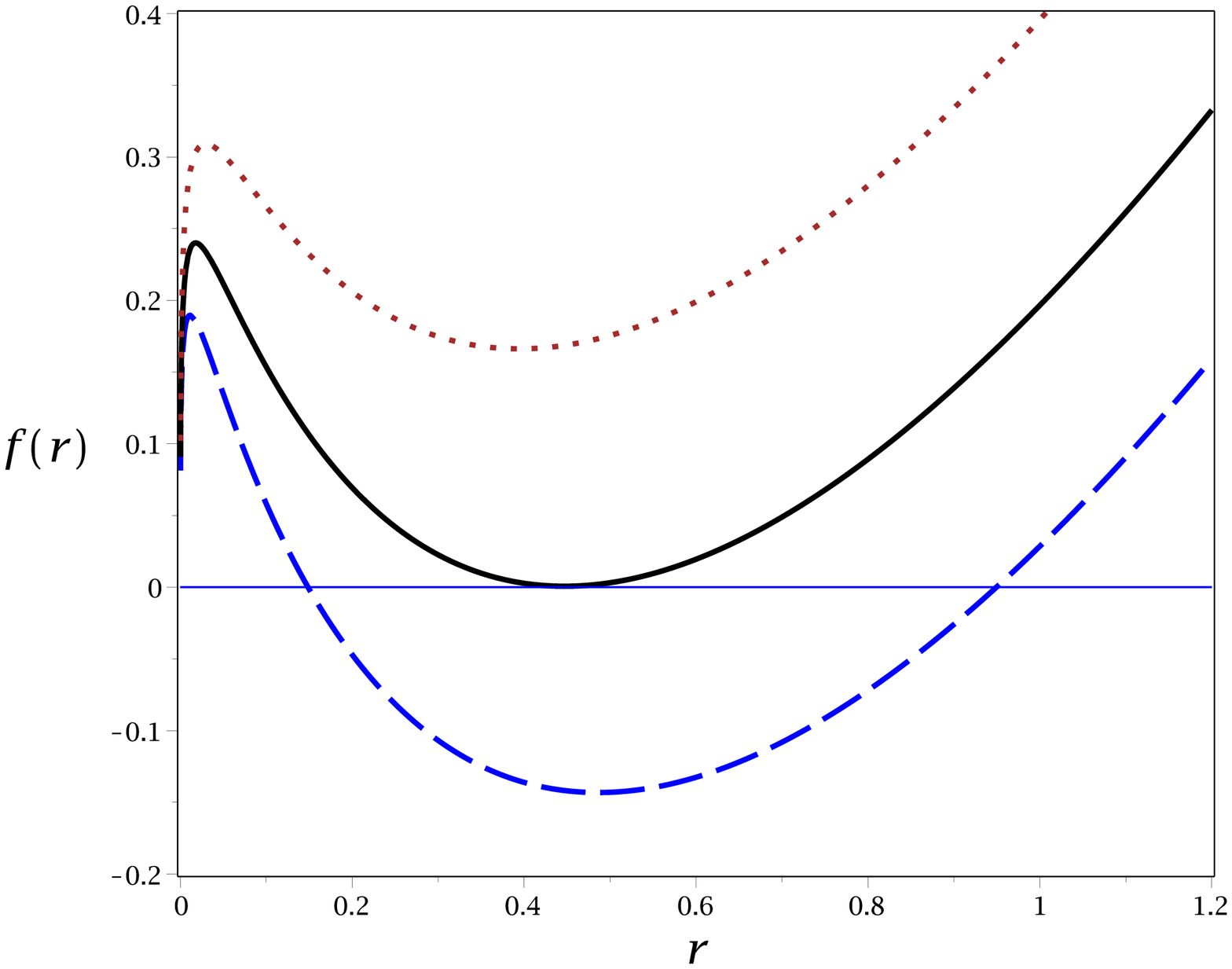} & \epsfxsize=7cm %
\epsffile{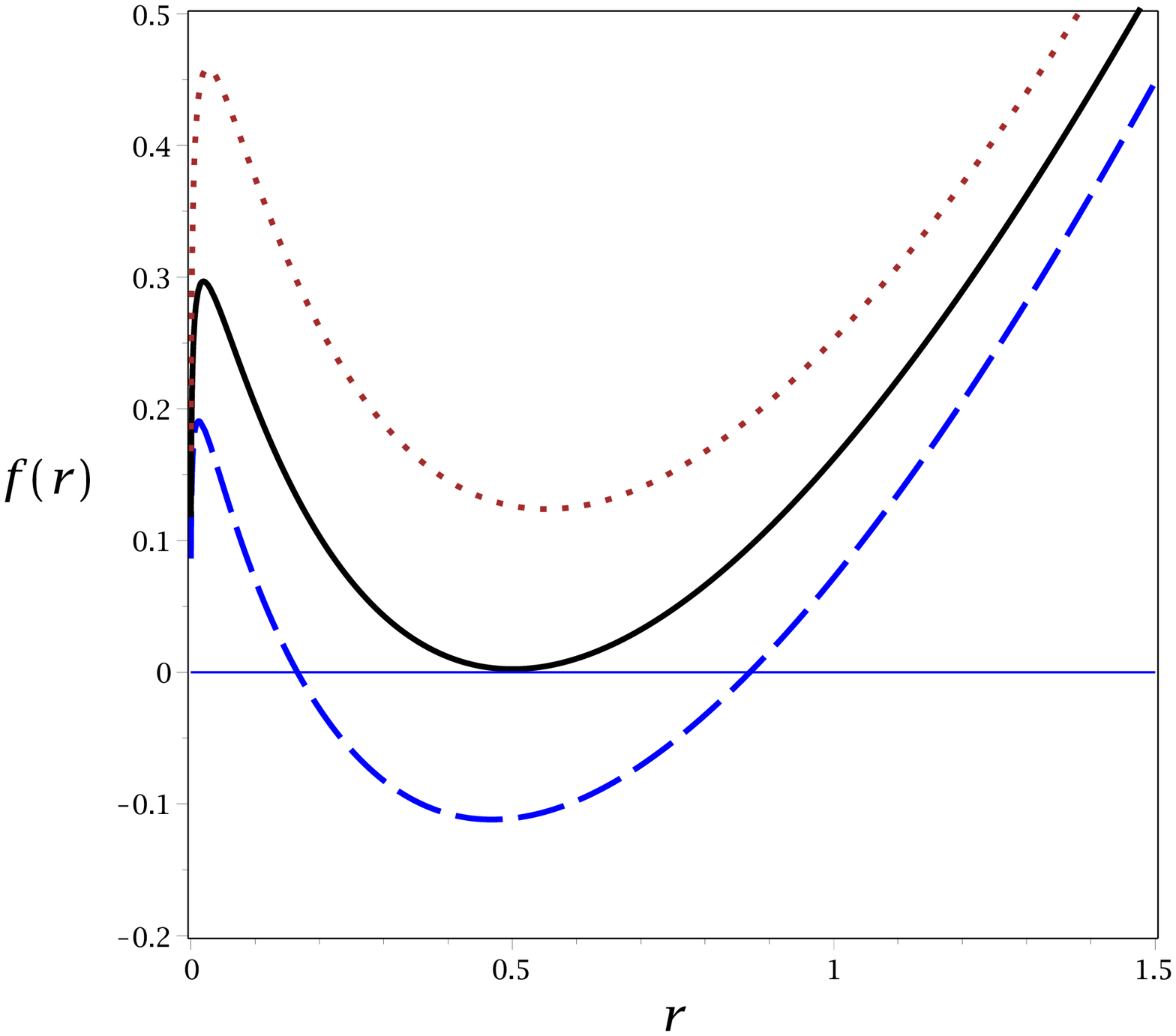} \\
\epsfxsize=7cm \epsffile{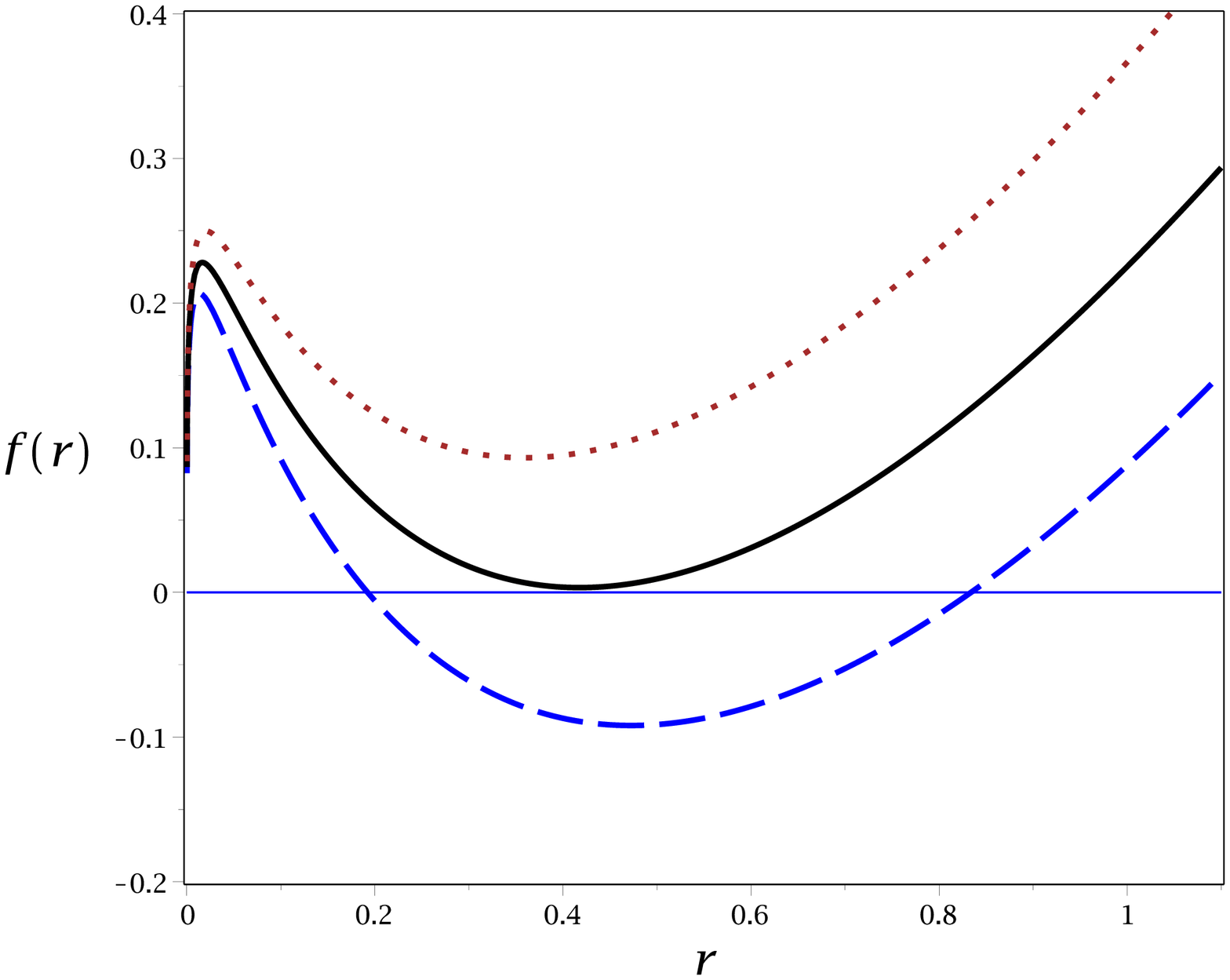} & \epsfxsize=7cm %
\epsffile{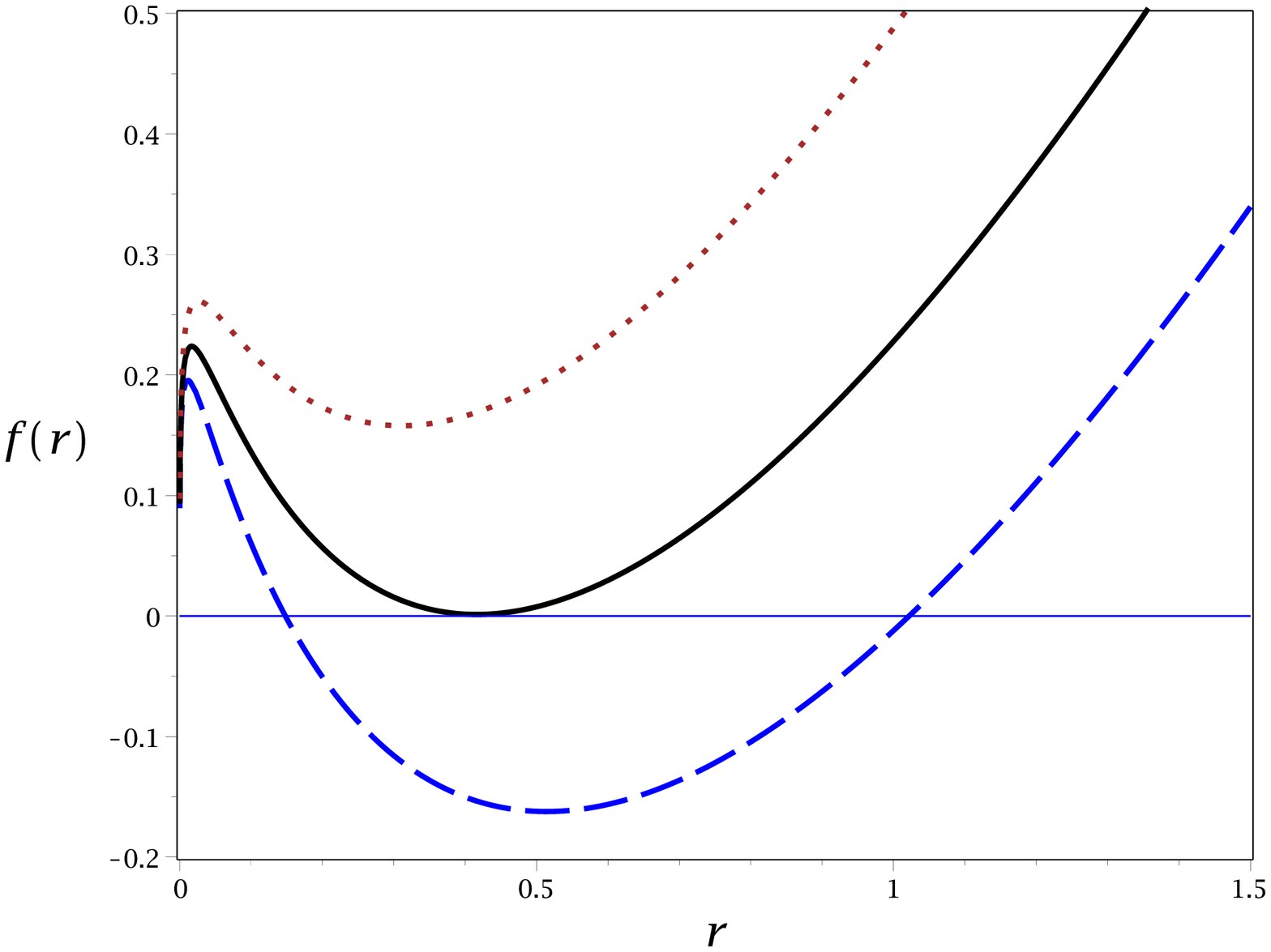}%
\end{array}
$%
\caption{$f(r)$ versus $r$, for $b=0.3$, $\Lambda =-1$.\newline
Left up panel: for $m=4$, $\protect\alpha =1$, $\protect\beta=0.5$, $q=0.490$
(dashed line), $q=0.518$ (continuous line) and $q=0.550$ (dotted line).%
\newline
Right up panel: for $m=4$, $\protect\alpha =1$, $q=0.5$, $\protect\beta %
=0.480$ (dashed line), $\protect\beta =0.615$ (continuous line) and $\protect%
\beta =0.850$ (dotted line).\newline
Right down panel: for $m=4$, $q=0.5$, $\protect\beta =0.5$, $\protect\alpha %
=1.000$ (dashed line), $\protect\alpha =1.021$ (continuous line) and $%
\protect\alpha =1.040$ (dotted line).\newline
Left down panel: for $q=0.5$, $\protect\beta =0.5$, $\protect\alpha =1$, $%
m=4.10$ (dashed line), $m=3.86$ (continuous line) and $m=3.60$ (dotted
line). }
\label{Fig2}
\end{figure}

The equation (\ref{RR0}) confirms that there is an essential singularity
located at $r=0$, while the equation. (\ref{RRinf}) shows that in the
presence dilaton field ($\alpha \neq 0$), the asymptotic behavior of the
solutions is not that of (A)dS.\ It is notable that, in the absence of
dilaton field ($\alpha =0$), the asymptotic behavior of the solutions is
(A)dS. We plot the obtained metric function Eq. (\ref{DilatonBTZBI}) in
figure \ref{Fig2}. This figure shows that the metric function may contain
real positive roots, and so, the curvature singularity can be covered with
an event horizon and interpreted as a black hole.

%%%%%%%%%%%%%%%%%%%%%%%%%%%%%%%%%%%%%%%%%%%%%%%%%%%%%%%%%%%%%%%
%%%%%%%%%%%%%%%%%%%%%%%%%%%%%%%%%%%%%%%%%%%%%%%%%%%%%%%%%%%%%%%

\subsection{Thermodynamical properties}

We can obtain the Hawking temperature by using Eq. (\ref{T}) for the this
black hole as
\begin{equation}
T=\frac{\left( \alpha ^{2}+1\right) r_{+}}{2\pi }\left( \frac{b}{r_{+}}%
\right) ^{2\gamma }\left[ 2\beta ^{2}\left( 1-H_{1_{+}}\right) -\Lambda %
\right] +\frac{q^{2}\left( \alpha ^{2}+1\right) }{\alpha ^{2}\pi r_{+}}\left[
H_{3}-\alpha ^{2}{}H_{2_{+}}+\frac{\alpha ^{2}\Gamma _{+}{}H_{4}}{\left(
\alpha ^{2}+2\right) }\right] .  \label{TBI}
\end{equation}
Notice that $H_{1_{+}}=H_{1}\left\vert _{r=r_{+}}\right.$ and $%
H_{2_{+}}=H_{2}\left\vert _{r=r_{+}}\right. $, and also that $H_{3}$ and $%
H_{4}$ are given by
\begin{eqnarray}
H_{3} &=&{}_2F_{1}\left( \left[ \frac{3}{2},\frac{\alpha ^{2}}{2}\right] ,%
\left[ \frac{\alpha ^{2}+2}{2}\right] ,-\Gamma _{+}\right) , \\
&&  \notag \\
H_{4} &=&{}_2F_{1}\left( \left[ \frac{3}{2},\frac{\alpha ^{2}+2}{2}\right] ,%
\left[ \frac{\alpha ^{2}+4}{2}\right] ,-\Gamma _{+}\right) ,
\end{eqnarray}%
where $\Gamma _{+}=\Gamma \left\vert _{r=r_{+}}\right. $. The electric
potential $U$ is obtained
\begin{equation}
U=\frac{q}{\gamma }\left( \frac{b}{r_{+}}\right) ^{\gamma }H_{2_{+}}.
\label{ElecPoPMI1}
\end{equation}
Using the area law of entropy in Einstein gravity, and also calculating the
flux of electromagnetic field at infinity we can obtain the entropy and the
electric charge of this black hole as%
\begin{eqnarray}
S &=&\frac{\pi r_{+}}{2}\left( \frac{b}{r_{+}}\right) ^{\gamma },
\label{entropyBI} \\
&&  \notag \\
Q &=&\frac{q}{2}.  \label{QBI}
\end{eqnarray}

According to the mentioned method for calculation of the total finite mass
of metric presented in Eq. (\ref{MetricI}), we can obtain the total mass as
\begin{equation}
M=\frac{m}{8}\left( 1-\gamma \right) b^{\gamma },  \label{massBI}
\end{equation}
which does not depend on the nonlinearity and on the electromagnetic field
directly since both the nonlinearity and the electromagnetic field vanish
for $r \rightarrow \infty$. Following the steps of the pervious sections,
the total mass of the black hole solution is obtained by evaluating the
metric function on its largest root
\begin{equation}
M(r_{+})=\frac{\left( \alpha ^{2}+1\right) }{2\left( \alpha ^{2}-2\right) }%
\left( \frac{b}{r_{+}}\right) ^{\gamma }\left\{ \left( \frac{\Lambda }{2}%
-\beta ^{2}\right) r_{+}^{2}+\beta ^{2}\left[ r_{+}^{2}\left( \frac{b}{r_{+}}%
\right) ^{2\gamma }H_{1_{+}}+\frac{q^{2}\left( \alpha ^{2}-2\right) H_{2_{+}}%
}{\alpha ^{2}\beta ^{2}}\right] \right\} ,  \label{MBI}
\end{equation}%
yielding as well to a Smarr-like formula given by
\begin{equation}
M(S,Q)=\frac{\left( \alpha ^{2}+1\right) }{2\left( \alpha ^{2}-2\right) }%
\left( \frac{\pi b}{2S}\right) ^{\alpha ^{2}}\left\{ \left( \frac{\Lambda }{2%
}-\beta ^{2}\right) \frac{\left( \frac{2S}{\pi }\right) ^{2\left( \alpha
^{2}+1\right) }}{b^{2\alpha ^{2}}}+\beta ^{2}\left[ \left( \frac{2S}{\pi}%
\right) ^{2} H_{1_{SQ}} +\frac{4Q^{2}\left( \alpha ^{2}-2\right) H_{2_{SQ}}}{%
\alpha ^{2}\beta ^{2}}\right] \right\} .  \label{SmarrBI}
\end{equation}%
where $H_{1_{SQ}}=H_{1_{+}}\left\vert _{r_{+}=b\left( \frac{2S}{\pi b}%
\right) ^{\alpha ^{2}+1},~q=2Q}~\right. $\ and $H_{2_{SQ}}=H_{2_{+}}\left%
\vert _{r_{+}=b\left( \frac{2S}{\pi b}\right) ^{\alpha
^{2}+1},~q=2Q}~\right. $. As a direct consequence, the first law of
thermodynamics holds
\begin{equation}
dM=TdS+UdQ,
\end{equation}%
with%
\begin{equation}
dM(S,Q)=\left( \frac{\partial M(S,Q)}{\partial S}\right) _{Q}dS+\left( \frac{%
\partial M(S,Q)}{\partial Q}\right) _{S}dQ,
\end{equation}%
and%
\begin{equation}
T=\left( \frac{\partial M}{\partial S}\right) _{Q}\ \ \ \ \&\ \ \ \ U=\left(
\frac{\partial M}{\partial Q}\right) _{S}.
\end{equation}

\subsection{Thermodynamic behavior}

In this section, we would like to stress how the Born-Infeld theory can
modify the thermodynamic behavior of the black holes. Our main motivation is
to distinguish the differences between the Maxwell and Born-Infeld theories
in the thermodynamic context of charged BTZ-dilatonic black holes.

\subsubsection{Mass/Internal energy}

\begin{figure}[tbp]
$%
\begin{array}{ccc}
\epsfxsize=5.5cm \epsffile{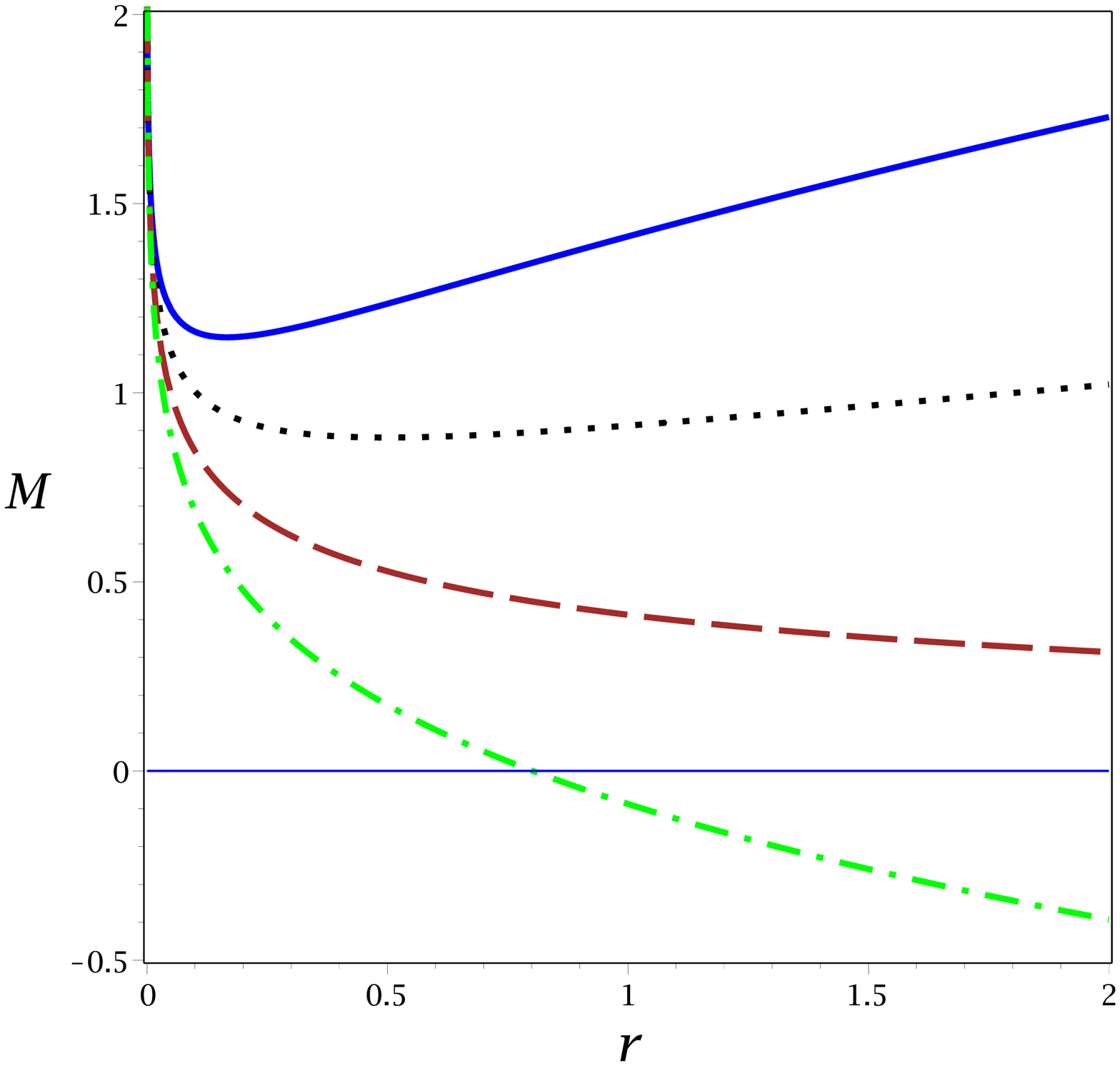} & \epsfxsize=5.5cm %
\epsffile{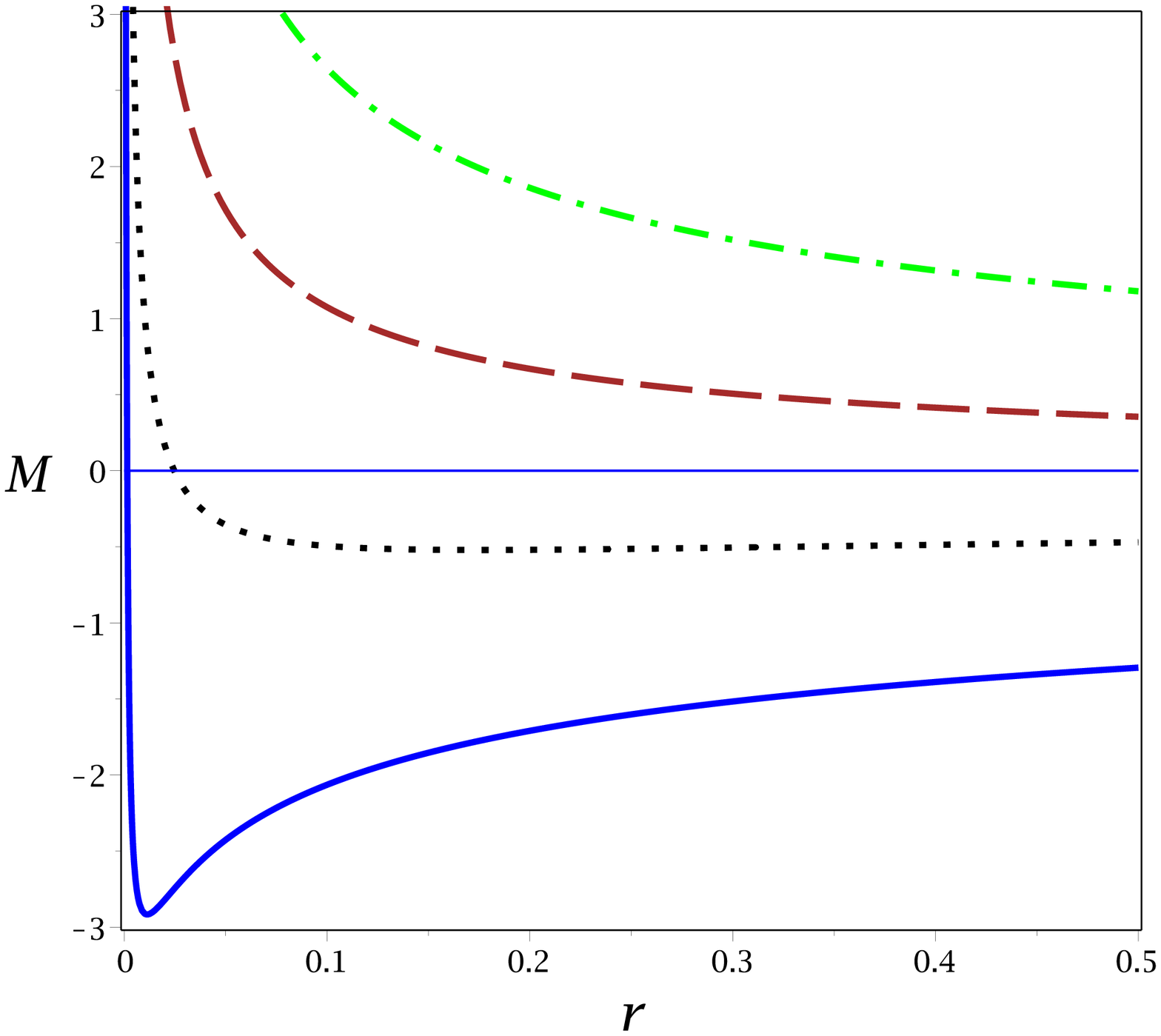} & \epsfxsize=5.5cm \epsffile{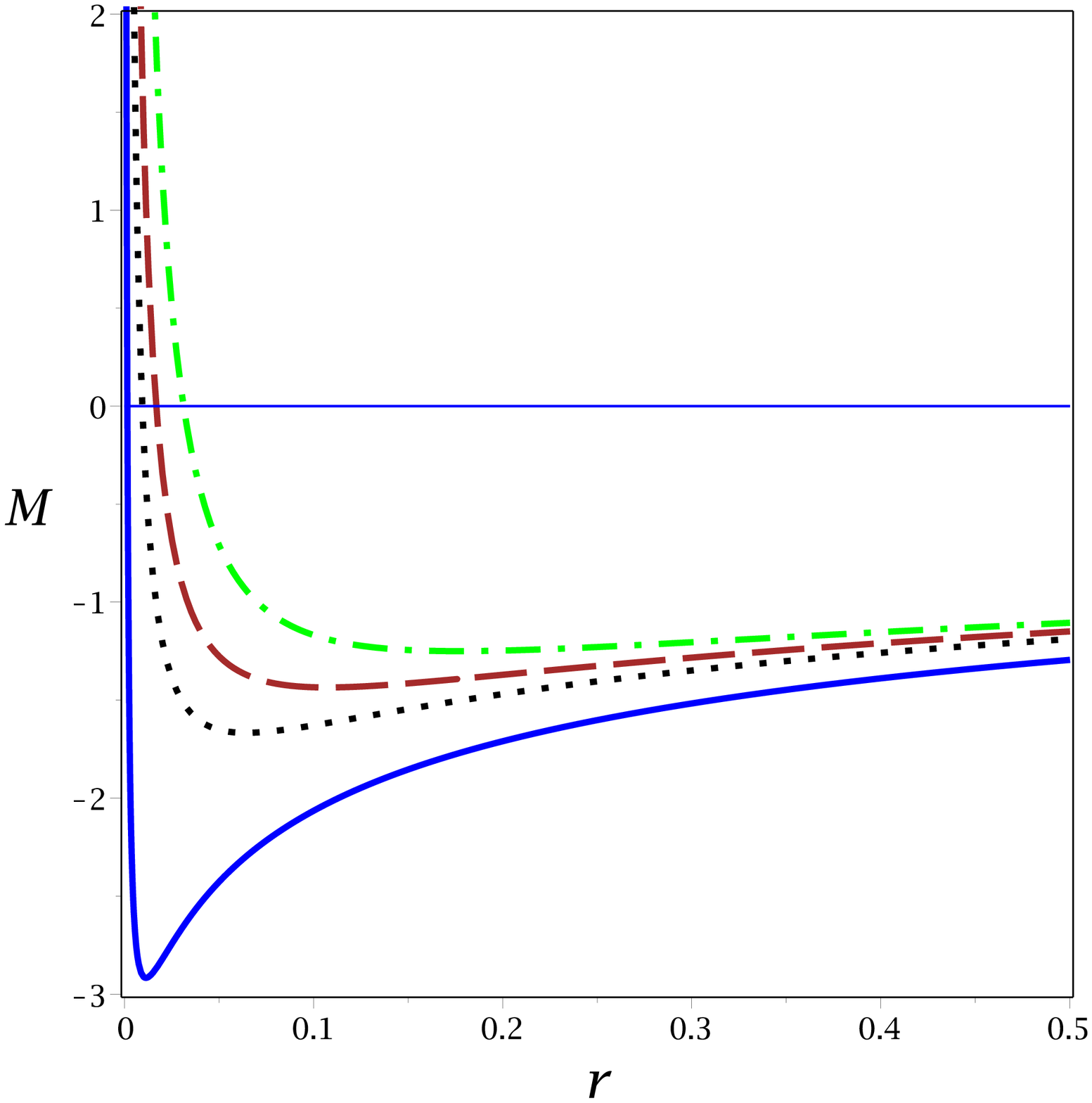}%
\end{array}
$%
\caption{$M$ versus $r_{+}$, for $b=1$, $q=1$, $\protect\beta=0.5$, $%
\Lambda=-2$ (continuous line), $\Lambda=-1 $ (dotted line), $\Lambda=0$
(dashed line) and $\Lambda=1$ (dashed-dotted line). \newline
Left panels: $\protect\alpha =1$; Middle panels: $\protect\alpha =2$.
\newline
Right panels: for $\Lambda=-2$, $\protect\alpha =2$, $\protect\beta=0.5$
(continuous line), $\protect\beta=1 $ (dotted line), $\protect\beta=1.5$
(dashed line) and $Maxwell~ case: \protect\beta \rightarrow \infty$
(dashed-dotted line). }
\label{FigBIA}
\end{figure}

As in the Maxwell case, the condition $\alpha\not=\sqrt{2}$ must also be
taken into consideration. This is not surprising since this condition is
only originated from the dilatonic part of the action. Interestingly enough,
one can see from the expression of the mass that for
\begin{equation}
\Lambda =2\beta ^{2},
\end{equation}%
the effects of the $\Lambda-$term are canceled by the nonlinearity term and
since the other two terms ($q$ and other $\beta $ terms) are positive
valued, the mass is positive everywhere without any root for this case. This
is one of the most important contributions of the generalization to BI field
which is not seen in the context of Maxwell case. Due to the complexity of
the mass relation, it is not possible to obtain the root of the mass
analytically. Therefore, we employ some numerical method and plot the
following diagrams (see left and middle panels of Fig. \ref{FigBIA}). It
could be seen that similar behaviors to those in Maxwell case are observed
here too. This means that the mass of these black holes could enjoy the
existences of root and two regions of positivity and negativity or it could
be positive valued everywhere. In case of the existence of root, the mass is
only positive valued before the root. The place of this root is a function
of the nonlinearity parameter (see right panel of Fig. \ref{FigBIA}). The
term $"\alpha ^{2}-2" $ is present in all terms except $q-$term. Therefore,
we can separate the effects of different terms (except the $q-$term) in the
mass into two categories: $\alpha >\sqrt{2}$ and $\alpha <\sqrt{2}$. We give
the details for $\alpha >\sqrt{2}$ since the opposite holds for the case $%
\alpha <\sqrt{2}$ (except for the $q-$term). For $\alpha >\sqrt{2}$, the $q-$%
term and one of the $\beta-$terms (the one which is coupled with the
electric charge) have positive contributions on the total value of the mass.
Whereas, the other $\beta-$term has always negative contribution. The
effects of the $\Lambda-$term depends on the choices of $\Lambda$ itself.
For negative $\Lambda$, the effect of this term is toward decreasing mass
while the opposite is observed for positive $\Lambda$. The existence of a
root for positive $\Lambda$ depends on following condition
\begin{equation*}
\Lambda <2\beta ^{2}.
\end{equation*}
If the mentioned condition is satisfied, it is possible to find a root for
the obtained mass, otherwise, the mass is always positive valued without any
root. The situation for a negative $\Lambda-$term is different and depends
on the following condition
\begin{equation*}
\left( \frac{\Lambda }{2}-\beta ^{2}\right) r_{+}^{2}>\beta ^{2}\left[
r_{+}^{2}\left( \frac{b}{r_{+}}\right) ^{2\gamma }H_{1_{+}}+\frac{%
q^{2}\left( \alpha ^{2}-2\right) H_{2_{+}}}{\alpha ^{2}\beta ^{2}}\right] ,
\end{equation*}%
which highly depends on choices of the nonlinearity parameter, $\beta$.

Here, we see that the effects of the nonlinearity parameter, hence the BI
generalization on the properties of the mass are significant. The presence
of $\beta $ provided an extra degree of freedom, and because of that the
mass of the solutions may have different behaviors. In the next sections, we
will give further details regarding this matter.
%%%%%%%%%%%%%%%%%%%%%%%%%%%%%%%%%%%%%%%%%%%%%%%%%%%%%%%%%%%%%%%
%%%%%%%%%%%%%%%%%%%%%%%%%%%%%%%%%%%%%%%%%%%%%%%%%%%%%%%%%%%%%%%

\subsubsection{Temperature}

%Temperature for these black holes contains several terms. I) Three
%$q-$terms in which, two of them have positive contributions on the
%values of temperature while the other one has negative effect. II)
%Two $\beta-$terms in which one of them contains electric charge
%parameter, while the other one has only dilatonic parameter. The
%earlier $\beta-$term has negative effect on the values of
%temperature while the later one has opposite (positive) effect. III)
%$\Lambda-$term in which if $\Lambda $ is negative, the effects of
%this term is toward increasing the temperature and the opposite
%exists for the positive values of $\Lambda$. Here too, the factors
%of $\Lambda $ and one of the $\beta-$terms are the same and by
%tuning the nonlinearity parameter properly ($\Lambda =2\beta ^{2}$),
%the effects of $\Lambda-$term could be canceled. This highlights one
%of the important effects of the generalization to BI theory.

\begin{figure}[tbp]
$%
\begin{array}{ccc}
\epsfxsize=5.5cm \epsffile{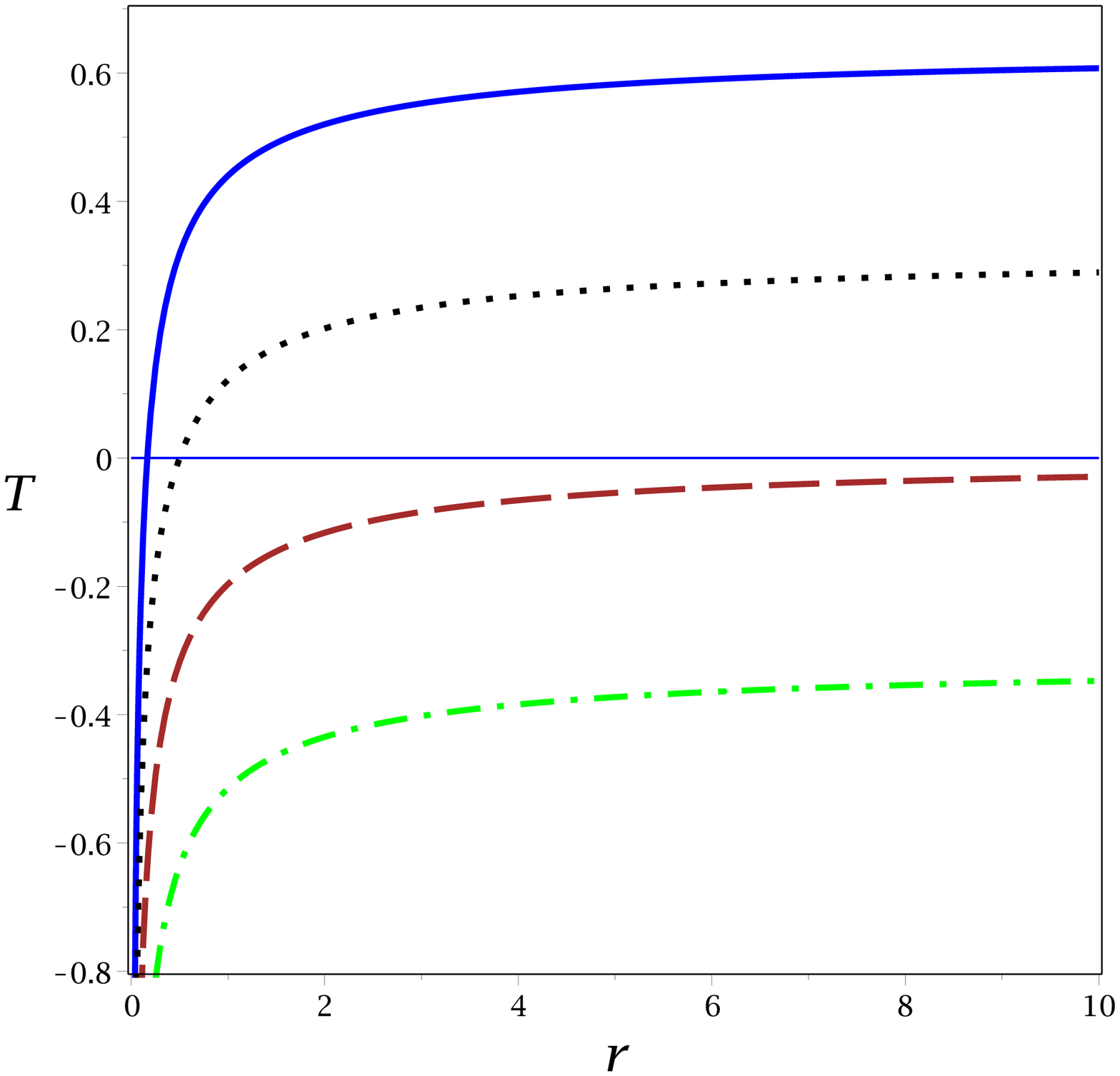} & \epsfxsize=5.5cm %
\epsffile{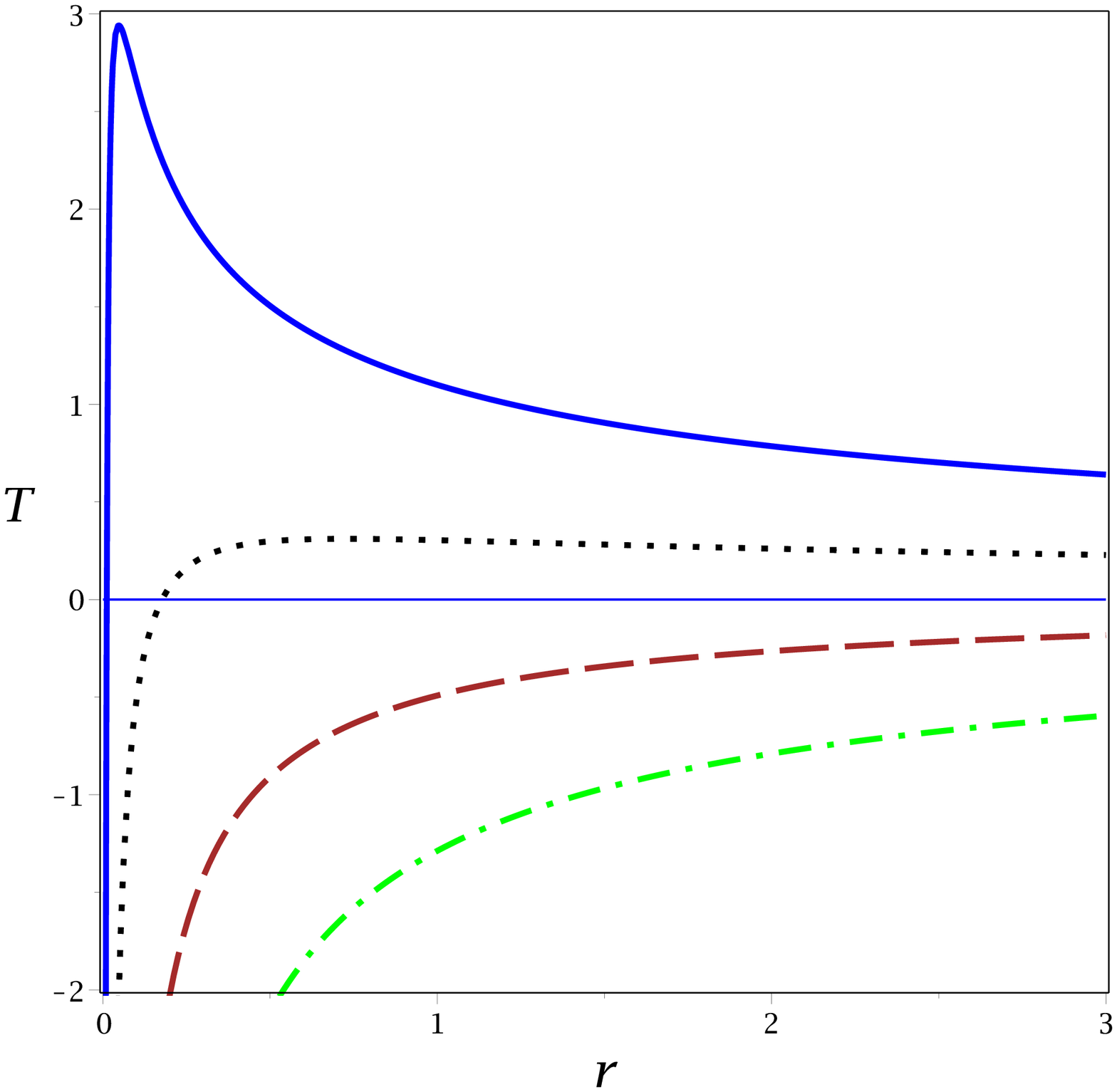} & \epsfxsize=5.5cm \epsffile{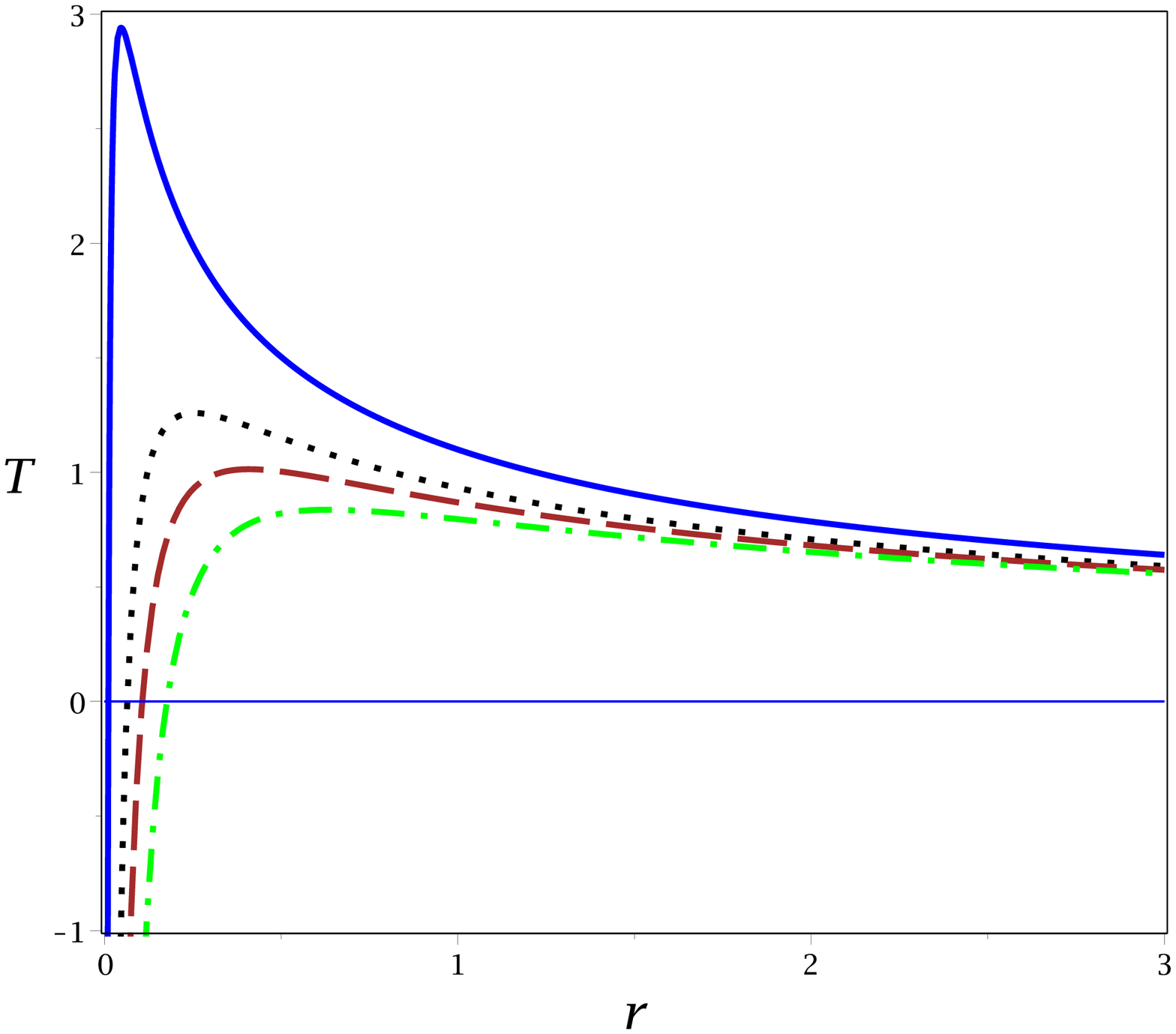}%
\end{array}
$%
\caption{$T$ versus $r_{+}$, for $b=1$, $q=1$, $\protect\beta =0.5$, $%
\Lambda =-2$ (continuous line), $\Lambda =-1$ (dotted line), $\Lambda =0$
(dashed line) and $\Lambda =1$ (dashed-dotted line). \newline
Left panels: $\protect\alpha =1$; Middle panels: $\protect\alpha =2$.
\newline
Right panels: for $\Lambda=-2$, $\protect\alpha =2$, $\protect\beta=0.5$
(continuous line), $\protect\beta=1 $ (dotted line), $\protect\beta=1.5$
(dashed line) and $Maxwell~ case: \protect\beta \rightarrow \infty$
(dashed-dotted line). }
\label{FigBI2}
\end{figure}

Obtaining the root and extremum point of temperature is not possible
analytically. As before, we will use some numerical method and we plot the
following diagrams (see left and middle panels of Fig. \ref{FigBI2}).
Evidently, depending on the choices of different parameters: I) The
temperature could be completely negative which in turn implies that the
solutions are not physical. II) The temperature could have one root and
before it, the temperature is negative valued. III) Having one root and one
maximum in which the maximum is located after the root and only after root
the temperature is positive. The places of the root and extremum depend on
the choices of the nonlinearity parameter (see right panel of Fig. \ref%
{FigBI2}). The generalization to nonlinear electromagnetic field provided us
with extra terms in the temperature which eventually modified the root, the
regions of negativity/positivity and the extremum of temperature. In
addition, this generalization results into one more degree of freedom which
could be used to tune out the effects of some part of the dilaton gravity.
This option was not possible in the context of Maxwell theory.

%%%%%%%%%%%%%%%%%%%%%%%%%%%%%%%%%%%%%%%%%%%%%%%%%%%%%%%%%%%%%%%
%%%%%%%%%%%%%%%%%%%%%%%%%%%%%%%%%%%%%%%%%%%%%%%%%%%%%%%%%%%%%%%

\subsubsection{Heat Capacity}

Our final study in this section is devoted to the heat capacity of the
nonlinearly charged solutions. By taking a closer look at Eq. (\ref{Heat}),
one can see that the heat capacity contains temperature and the derivations
of entropy and temperature with respect to the horizon radius. The
conditions regarding the roots and positivity/negativity of the temperature
were given in the last section. The derivation of the entropy with respect
to the horizon radius does not produce any singular point and it does not
contain terms which could contribute to negativity/positivty of the heat
capacity. Therefore, we focus our attention on $\left( \frac{\partial T}{%
\partial r_{+}}\right) _{Q}$ given by
\begin{eqnarray}
\left( \frac{\partial T}{\partial r_{+}}\right) _{Q} &=&\frac{\left( \alpha
^{2}-1\right) \left( \frac{\Lambda }{2}-\beta ^{2}\right) \left( \frac{b}{%
r_{+}}\right) ^{2\gamma }}{\pi }+\frac{\left( \alpha ^{2}-1\right) \beta
^{2}\left( \frac{b}{r_{+}}\right) ^{2\gamma }}{\pi }H_{1_{+}}-\frac{\left(
2\alpha ^{2}-1\right) H_{3_{+}}-\alpha ^{2}\left( \alpha ^{2}+1\right)
H_{2_{+}}}{\pi \alpha ^{2}r_{+}^{2}} \\
&&+\frac{\left( 2\alpha ^{2}+3\right) H_{4}-3H_{5}}{\pi \beta ^{2}\left(
\alpha ^{2}+2\right) r_{+}^{4}\left( \frac{b}{r_{+}}\right) ^{2\gamma }}+%
\frac{3q^{6}{}H_{6}}{\pi \beta ^{4}\left( \alpha ^{2}+4\right)
r_{+}^{6}\left( \frac{b}{r_{+}}\right) ^{4\gamma }},
\end{eqnarray}
where $H_{5}$ and $H_{6}$ are defined as follows
\begin{eqnarray}
H_{5} &=&{}_2F_{1}\left( \left[ \frac{5}{2},\frac{\alpha ^{2}+2}{2}\right] ,%
\left[ \frac{\alpha ^{2}+4}{2}\right] ,-\Gamma _{+}\right) , \\
&&  \notag \\
H_{6} &=&{}_2F_{1}\left( \left[ \frac{5}{2},\frac{\alpha ^{2}+4}{2}\right] ,%
\left[ \frac{\alpha ^{2}+6}{2}\right] ,-\Gamma _{+}\right) .
\end{eqnarray}

\begin{figure}[tbp]
$%
\begin{array}{ccc}
\epsfxsize=5.5cm \epsffile{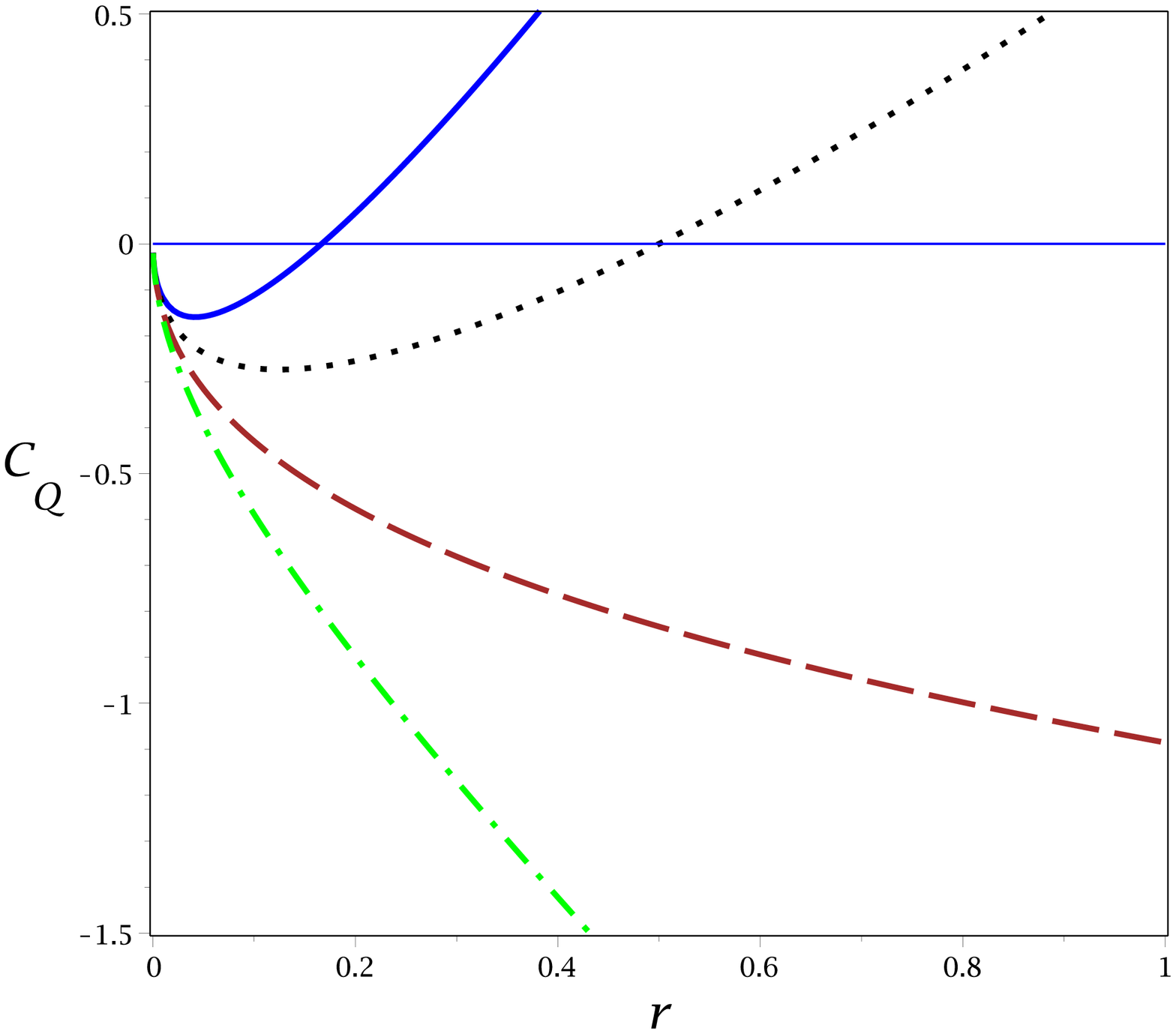} & \epsfxsize=5.5cm %
\epsffile{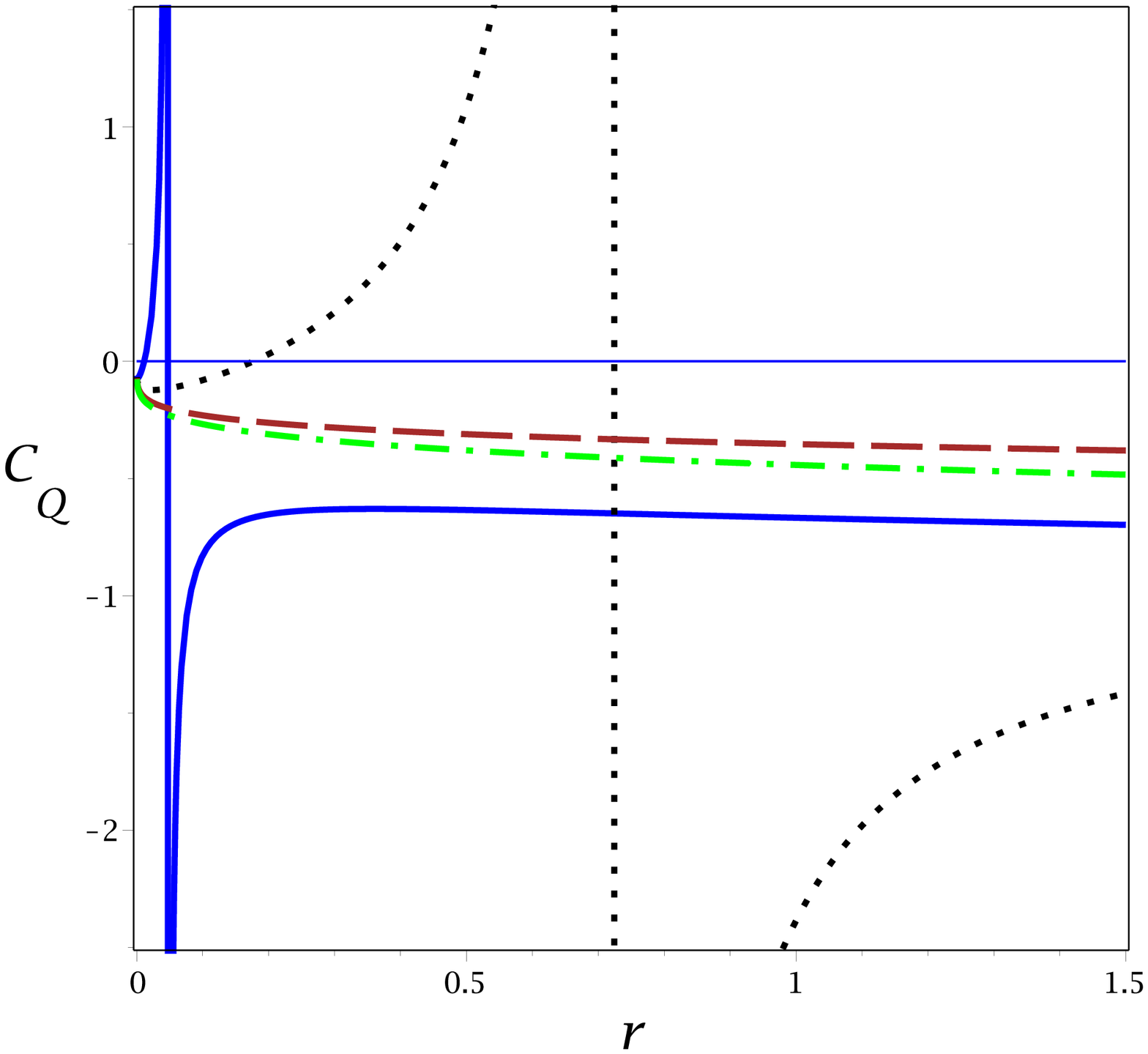} & \epsfxsize=5.5cm \epsffile{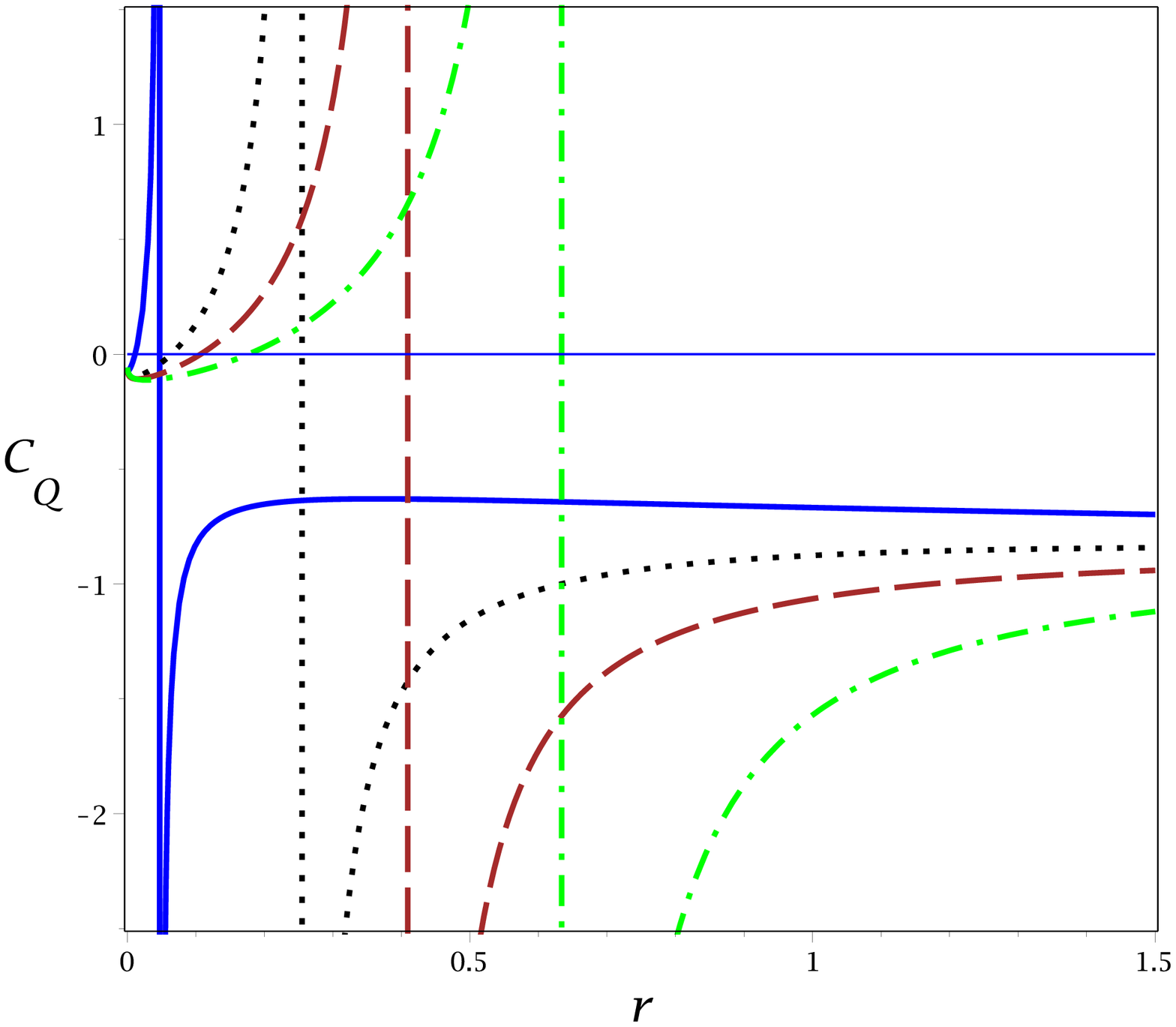}%
\end{array}
$%
\caption{$C_{Q}$ versus $r_{+}$, for $b=1$, $q=1$, $\protect\beta=0.5$, $%
\Lambda=-2$ (continuous line), $\Lambda=-1 $ (dotted line), $\Lambda=0$
(dashed line) and $\Lambda=1$ (dashed-dotted line). \newline
Left panels: $\protect\alpha =1$; Right panels: $\protect\alpha =2$. \newline
Right panels: for $\Lambda=-2$, $\protect\alpha =2$, $\protect\beta=0.5$
(continuous line), $\protect\beta=1 $ (dotted line), $\protect\beta=1.5$
(dashed line) and $Maxwell~ case: \protect\beta \rightarrow \infty$
(dashed-dotted line). }
\label{FigBI3}
\end{figure}

Similar to other quantities, here, we are able to tune out the effects of $%
\Lambda-$term by suitable choices of the nonlinearity parameter. In other
words, it is possible to cancel out the effects of $\Lambda$ in the heat
capacity of BI solutions by choosing $\Lambda=2\beta^{2}$. Once more we
point it out that such possibility is present in BI generalization while it
is not seen in the linear Maxwell theory. The presence of nonlinear
electromagnetic field provided a complicated system of terms in the heat
capacity. Unfortunately, such complication does not allow us to extract
divergence points of the heat capacity analytically. We have plotted the
following diagrams (see left and middle panels of Fig. \ref{FigBI3}).
Depending on the choices of different parameters, one of the following cases
would take place for the heat capacity; I) Two states of stable and unstable
which are separated by a root. In the previous section, it was shown that
the root of the temperature, hence the heat capacity, is a function of
nonlinearity parameter. II) The heat capacity could be negative everywhere
without any root or divergency. In this case the solutions are unstable but
according to the results of previous section, the temperature is also
negative which indicates that the solutions are not physical. III) The heat
capacity could enjoy one root and one divergency. The divergency points out
the existence of a phase transition. Before the root, the heat capacity is
also negative. Therefore, the only physically stable solutions exist between
the root and divergency of the heat capacity. The plotted diagram for the
variation of the nonlinearity parameter (see right panel of Fig. \ref{FigBI3}%
), $\beta $, shows that the location of divergency is a function of this
parameter. This indicates that the region in which physical stable solutions
exist, depends on the choices of $\beta$. This highlights another
significant effect of the nonlinearity parameter.

%%%%%%%%%%%%%%%%%%%%%%%%%%%%%%%%%%%%%%%%%%%%%%%%%%%%%%%%%%%%%%%
%%%%%%%%%%%%%%%%%%%%%%%%%%%%%%%%%%%%%%%%%%%%%%%%%%%%%%%%%%%%%%%

\section{Conclusion}

The paper at hand regarded BTZ black holes in the presence of two
generalizations which are motivated by string theory: dilaton gravity and
Born-Infeld nonlinear electromagnetic field.

First, the solutions in the presence of dilaton gravity were extracted and
their thermodynamical properties were studied. It was shown that in
comparison to the Einstein gravity, here, the mass of these black holes
could enjoy the existence of root and a region of negative mass/internal
energy. In addition, specific limits for the dilaton parameter and $\Lambda$
were obtained and it was shown that thermodynamically speaking, only for
specific region of dilaton parameter, physical solutions exist. Therefore,
although generalization to dilaton gravity provided us with new properties
for the solutions, at the same time, it imposed specific limits on them as
well. In other words, introducing new properties into solutions by dilatonic
generalization was obtained at the cost of harder restrictions on the
solutions.

Next, Born-Infeld generalization was implied to the action. It was shown
that this generalization provides the possibility of canceling a part of
dilatonic contribution by suitable choices of parameters. In addition, it
was shown that although some of the Maxwell conditions for having
thermodynamically physical solutions stand for this case as well, the
general behavior of the solutions including phase transition point, region
of stable solutions, conditions of having physical solutions were modified
due to the contributions of BI theory.

The study conducted here could be employed to investigate aspects
such as superconductor properties, holographical principles and
entropy spectrum. Specially, it is interesting to study the
central charges of $1+1$ theories and understand the effects of
the dilatonic gravity and BI generalization in this context. In
addition, as it is known, we can discuss dyon solutions of our
dilatonic setup with two horizons. We address these subjects in
the forthcoming works.
\begin{acknowledgements}
We thank Shiraz University Research Council. This work has been
supported financially by the Research Institute for Astronomy and
Astrophysics of Maragha, Iran.
\end{acknowledgements}

%%%%%%%%%%%%%%%%%%%%%%%%%%%%%%%%%%%%%%%%%%%%%%%%%%%%%%%%%%%%%%%%%%%%%%%%%%%%%%%%%%%%%%%%%%%%%%%


\begin{thebibliography}{99}
\bibitem{MignemiW} S. Mignemi and D. L. Wiltshire, Phys. Rev. D \textbf{46},
1475 (1992).

\bibitem{PolettiTW} S. J. Poletti, J. Twamley and D. L. Wiltshire, Phys.
Rev. D \textbf{51}, 5720 (1995).

\bibitem{Cai} R. G. Cai, J. Y. Ji and K. S. Soh, Phys. Rev D \textbf{57},
6547 (1998).

\bibitem{LiouvilleI} M. Kord Zangeneh, A. Sheykhi and M. H. Dehghani, Phys.
Rev. D \textbf{91}, 044035 (2015).

\bibitem{LiouvilleII} M. Kord Zangeneh, A. Sheykhi and M. H. Dehghani, Eur.
Phys. J. C \textbf{75}, 497 (2015).

\bibitem{LiouvilleIII} S. H. Hendi, B. Eslam Panah, S. Panahiyan and M.
Momennia, Eur. Phys. J. C \textbf{77}, 647 (2017).

\bibitem{LiouvilleIV} A. Sheykhi, S. H. Hendi, F. Naeimipour, S. Panahiyan
and B. Eslam Panah, Can. J. Phys. \textbf{94}, 1045 (2016).

\bibitem{GaoZh} C. J. Gao and S. N. Zhang, Phys. Rev. D \textbf{70}, 124019
(2004).

\bibitem{ShDH} A. Sheykhi, M. H. Dehghani and S. H. Hendi, Phys. Rev. D.
\textbf{81}, 084040 (2010).

\bibitem{KoikawaY} T. Koikawa and M. Yoshimura, Phys. Lett. B \textbf{189},
29 (1987).

\bibitem{GibbonsM} G. W. Gibbons and K. Maeda, Nucl. Phys. B \textbf{298},
741 (1988).

\bibitem{BrillH} D. Brill and J. Horowitz, Phys. Lett. B \textbf{262}, 437
(1991).

\bibitem{Tarrio:2011de} J.~Tarrio and S.~Vandoren, JHEP \textbf{09}, 017
(2011).

\bibitem{nonlinA} M.~Hassaine and C.~Martinez, Phys.\ Rev.\ D \textbf{75},
027502 (2007).

\bibitem{nonlinB} M.~Hassaine and C.~Martinez, Class.\ Quantum Gravit.\
\textbf{25}, 195023 (2008).

\bibitem{nonlinC} H.~Maeda, M.~Hassaine and C.~Martinez, Phys.\ Rev.\ D
\textbf{79}, 044012 (2009).

\bibitem{Zangeneh:2015uwa} M.~K.~Zangeneh, A.~Sheykhi and M.~H.~Dehghani,
Phys.\ Rev.\ D \textbf{92}, 024050 (2015).

\bibitem{Kord} M. Kord Zangeneh, A. Dehyadegari, M. R. Mehdizadeh, B. Wang
and A. Sheykhi, Eur. Phys. J. C \textbf{77}, 423 (2017).

\bibitem{Zangeneh1} M. Kord Zangeneh, A. Dehyadegari, A. Sheykhi and M. H.
Dehghani, JHEP \textbf{03}, 037 (2016).

\bibitem{Zangeneh2} A. Dehyadegari, A. Sheykhi and M. Kord Zangeneh, Phys.
Lett. B \textbf{758}, 226 (2016).

\bibitem{Zangeneh3} M. Kord Zangeneh, S. S. Hashemi, A. Dehyadegari, A.
Sheykhi and B. Wang, [arXiv:1710.10162].

\bibitem{HendiBEN} S. H. Hendi, G. H. Bordbar, B. Eslam Panah and M. Najafi,
Astrophys. Space Sci. \textbf{358}, 30 (2015).

\bibitem{HendiFEP} S. H. Hendi, M. Faizal, B. Eslam Panah and S. Panahiyan,
Eur. Phys. J. C \textbf{76}, 296 (2016).

\bibitem{HendiPTEP} S. H. Hendi, B. Eslam Panah and S. Panahiyan, Prog.
Theor. Exp. Phys. \textbf{2016}, 103A02 (2016).

\bibitem{BTZ1} M. Banados, C. Teitelboim and J. Zanelli, Phys. Rev. Lett.
\textbf{69}, 1849(1992).

\bibitem{BTZ2} M. Banados, M. Henneaux, C. Teitelboim and J. Zanelli, Phys.
Rev. D \textbf{48}, 1506 (1993).

\bibitem{Emparan} R. Emparan, G. T. Horowitz and R. C. Myers, JHEP \textbf{01%
}, 021 (2000).

\bibitem{ParsonsR} J. Parsons and S. F. Ross, JHEP \textbf{04}, 134 (2009).

\bibitem{MyungKM} Y. S. Myung, Y. W. Kim, T. Moon and Y. J. Park, Phys. Rev.
D \textbf{84}, 024044 (2011).

\bibitem{Frodden} E. Frodden, M. Geiller, K. Noui and A. Perez, JHEP \textbf{%
05}, 139 (2013).

\bibitem{Bravo} M. Bravo-Gaete and M. Hassaine, Phys. Rev. D \textbf{90},
024008 (2014).

\bibitem{HendiES} S. H. Hendi, B. Eslam Panah and R. Saffari, Int. J. Mod.
Phys. D \textbf{23} , 1450088 (2014).

\bibitem{WuLZ} X. H. Wu, R. Li and J. K. Zhao, Phys. Rev. D \textbf{93},
064008 (2016).

\bibitem{KordBTZ} Z. -Y. Tang, C. -Y. Zhang, M. Kord Zangeneh, B. Wang and
J. Saavedra, Eur. Phys. J. C 77, 390 (2017).

\bibitem{Witten07} E. Witten, [arXiv:0706.3359].

\bibitem{AnacletoB} M. A. Anacleto, F. A. Brito and E. Passos, Phys. Lett. B
\textbf{743}, 184 (2015).

\bibitem{Witten} E. Witten, Adv. Theor. Math. Phys. \textbf{2}, 505 (1998).

\bibitem{LarranagaI} A. Larranaga, Commun. Theor. Phys. \textbf{50}, 1341
(2008).

\bibitem{BTZnon} M. Cataldo and A. Garcia, Phys. Lett. B \textbf{456}, 28
(1999).

\bibitem{Yamazaki} R. Yamazaki and D. Ida, Phys. Rev. D \textbf{64}, 024009
(2001).

\bibitem{HendiEPJCthree} S. H. Hendi, B. Eslam Panah, M. Momennia and S.
Panahiyan, Eur. Phys. J. C \textbf{75}, 457 (2015).

\bibitem{BTZlikeI} S. H. Hendi, Eur. Phys. J. C \textbf{71}, 1551 (2011).

\bibitem{BTZlikeII} S. G. Ghosh, Int. J. Mod. Phys. D \textbf{21}, 1250022
(2012).

\bibitem{BTZmassive} S. H. Hendi, B. Eslam Panah and S. Panahiyan, JHEP
\textbf{05}, 029 (2016).

\bibitem{CM1A} K. C. K. Chan and R. B. Mann, Phys. Rev. D \textbf{50}, 6385
(1994).

\bibitem{CM1B} M. Dehghani, Phys. Rev. D \textbf{96}, 044014 (2017).

\bibitem{CM1C} S. H. Hendi, B. Eslam Panah, S. Panahiyan and A. Sheykhi,
Phys. Lett. B \textbf{767}, 214 (2017).

\bibitem{RainbowI} C. Leiva and I. Espinoza, [arXiv:1006.2406].

\bibitem{RainbowIII} S. H. Hendi, B. Eslam Panah, S. Panahiyan, Prog. Theor.
Exp. Phys. \textbf{2016}, 103A02 (2016).

\bibitem{NewmassiveI} Y. S. Myung, Y. W. Kim, T. Moon and Y. J. Park, Phys.
Rev. D \textbf{84}, 024044 (2011).

\bibitem{NewmassiveII} T. Moon and Y. S. Myung, Phys. Rev. D \textbf{88},
124014 (2013).

\bibitem{Lifshitz} E. Ayon-Beato, A. Garbarz, G. Giribet and M. Hassaine,
Phys. Rev. D \textbf{80}, 104029 (2009).

\bibitem{massrain} S. H. Hendi, S. Panahiyan, S. Upadhyay and B. Eslam
Panah, Phys. Rev. D \textbf{95}, 084036 (2017).

\bibitem{BTZgravI} V. Suneeta, R. K. Kaul and T. R. Govindarajan, Mod. Phys.
Lett. A \textbf{14}, 349 (1999).

\bibitem{BTZgravII} H. Saida and J. Soda, Phys. Lett. B \textbf{471}, 358
(2000).

\bibitem{BTZgravIII2} G. de Berredo-Peixoto, Class. Quantum Gravit. \textbf{%
20}, 3983 (2003).

\bibitem{BTZgravIII3} E. A. Bergshoeff, O. Hohm and P. K. Townsend, Phys.
Rev. Lett. \textbf{102}, 201301 \ (2009).

\bibitem{BTZgravVA} E. Frodden, M. Geiller, K. Noui and A. Perez, JHEP
\textbf{05}, 139 (2013)

\bibitem{BTZgravVB} G. G. L. Nashed, S. Capozziello, [arXiv:1710.06620].

\bibitem{BTZgravVC} D. V. Singh and S. Siwach, Class. Quantum Gravit.
\textbf{30}, 235034 (2013).

\bibitem{Myung2006} Y. S. Myung, Phys. Lett. B \textbf{638}, 515 (2006).

\bibitem{LiuPW} Y. Liu, Q. Pan and B. Wang, Phys. Lett. B \textbf{702}, 94
(2011).

\bibitem{Germani} C. Germani and G. P. Procopio, Phys. Rev. D \textbf{74},
044012 (2006).

\bibitem{FroddenGNP} E. Frodden, M. Geiller, K. Noui and A. Perez, JHEP
\textbf{05}, 139 (2013).

\bibitem{Caputa} P. Caputa, V. Jejjala and H. Soltanpanahi, Phys. Rev. D
\textbf{89}, 046006 (2014).

\bibitem{de la Fuente} A. de la Fuente and R. Sundrum, JHEP \textbf{09}, 073
(2014).

\bibitem{Chaturvedi} P. Chaturvedi and G. Sengupta, Phys. Rev. D \textbf{90}%
, 046002 (2014).

\bibitem{BI} M. Born and L. Infeld, Proc. Roy. Soc. Lond. \textbf{144}, 425
(1934).

\bibitem{BIStringI} E. S. Fradkin and A. A. Tseytlin, Phys. Lett. B \textbf{%
163}, 123 (1985).

\bibitem{BIStringII} D. L. Wiltshire, Phys. Rev. D \textbf{38}, 2445 (1988).

\bibitem{BIStringIII} R. G. Leigh, Mod. Phys. Lett. A \textbf{4}, 2767
(1989).

\bibitem{BIStringIV} M. Cataldo and A. Garcia, Phys. Lett. B \textbf{456},
28 (1999).

\bibitem{BIStringV} G. W. Gibbons and C. A. R. Herdeiro, Class. Quant. Grav.
\textbf{18}, 1677 (2001).

\bibitem{BIStringVI} G. W. Gibbons, Rev. Mex. Fis. \textbf{49S1}, 19 (2003).

\bibitem{Hoffmann} B. Hoffmann, Phys. Rev. \textbf{47}, 877 (1935).

\bibitem{BHBI1} M. Demianski, Found. Phys. \textbf{16}, 187 (1986).

\bibitem{BHBI2} H. P. de Oliveira, Class. Quant. Grav. \textbf{11}, 1469
(1994).

\bibitem{BHBI3} S. Fernando and D. Krug, Gen. Rel. Grav. \textbf{35}, 129
(2003).

\bibitem{BHBI4} R. -G. Cai, D. -W. Pang and A. Wang, Phys. Rev. D \textbf{70}%
, 124034 (2004).

\bibitem{BHBI5} T. K. Dey, Phys. Lett. B \textbf{595}, 484 (2004).

\bibitem{BHBI6} D. J. Cirilo Lombardo, Gen. Rel. Grav. \textbf{37}, 847
(2005).

\bibitem{BHBI7} Y. S. Myung, Y. -W. Kim and Y. -J. Park, Phys. Rev. D
\textbf{78}, 044020 (2008).

\bibitem{BHBI8} O. Miskovic and R. Olea, Phys. Rev. D \textbf{77}, 124048
(2008).

\bibitem{BHBI9} W. A. Chemissany, M. de Roo and S. Panda, Class. Quant.
Grav. \textbf{25}, 225009 (2008).

\bibitem{BHBI10} R. -G. Cai and Y. -W. Sun, JHEP \textbf{09}, 115 (2008).

\bibitem{BHBI11} P. Li, R. -H. Yue and D. -C. Zou, Commun. Theor. Phys.
\textbf{56}, 845 (2011).

\bibitem{BHBI12} R. Banerjee and D. Roychowdhury, Phys. Rev. D \textbf{85},
104043 (2012).

\bibitem{BHBI13} D. -C. Zou, S. -J. Zhang and B. Wang, Phys. Rev. D \textbf{%
89}, 044002 (2014).

\bibitem{BHBI14} X. -X. Zeng, X. -M. Liu and L. -F. Li, Eur. Phys. J. C
\textbf{76}, 616 (2016).

\bibitem{BHBI15} K. Meng and D. -B. Yang, [arXiv:1712.08798].

\bibitem{Ozer} M. Ozer and M. O. Taha, Phys. Rev. D \textbf{45}, 997 (1992).

\bibitem{Chan} K. C. K. Chan, J. H. Horne and R. B. Mann, Nucl. Phys. B
\textbf{447}, 441 (1995).

\bibitem{Yazadjiev} S. S. Yazadjiev, Class. Quantum Gravit. \textbf{22},
3875 (2005).

\bibitem{Cvetic1} M. Cvetic and S. S. Gubser, JHEP \textbf{04}, 024 (1999).

\bibitem{Bekenstein1} J. D. Bekenstein, Phys. Rev. D \textbf{7}, 2333 (1973).

\bibitem{Bekenstein2} J. M. Bardeen, B. Carter and S. W. Hawking, Commun.
Math. Phys. \textbf{31}, 161 (1973).

\bibitem{Bekenstein3} S.W. Hawking and C. J. Hunter, Phys. Rev. D \textbf{59}%
, 044025 (1999).

\bibitem{Mamasani} S. H. Hendi, S. Panahiyan and R. Mamasani, Gen. Relativ.
Gravit. \textbf{47}, 91 (2015).

\bibitem{Dehghani} M. H. Dehghani and N. Farhangkhah, Phys. Rev. D \textbf{71%
}, 044008 (2005).
\end{thebibliography}
\end{document}